\documentclass[11pt]{article}

\usepackage{amsmath,amsthm}
\usepackage{amssymb}
\usepackage{graphicx}
\usepackage{color}
\usepackage[latin1]{inputenc}
\usepackage[english]{babel}
\usepackage{a4wide}
\usepackage{algorithmic}
\usepackage{algorithm}
\usepackage{lscape}
\usepackage{graphicx,color}
\usepackage{array}
\usepackage{subcaption}
\usepackage{tikz}
\usetikzlibrary{shapes}
\usetikzlibrary{positioning}

\usepackage{lscape}
\usepackage{natbib}

\usepackage{tikz}
 
\newcommand{\myScale}{0.8} 
 
\newcommand{\myColor}{gray!40}

\newcommand{\drawThreeStacks}[6]{
	\fill[\myColor] (0 + #5,0) rectangle (1 + #5,#1);
	\fill[\myColor] (1 + #5,0) rectangle (2 + #5,#2);
	\fill[\myColor] (2 + #5,0) rectangle (3 + #5,#3);
	\draw (0 + #5,0) grid (3 + #5, #4 + 0.5);
	
	\draw (0 + #5,0) -- coordinate (x axis mid) (3 + #5,0);
    \draw (0 + #5,0) -- coordinate (y axis mid) (0 + #5,#4);
    
    \foreach \x in {1,2,3}{
     \draw (\x - 0.5 + #5, -0.4) node{\x};
    }
}

\begin{document}

\title{A Local-Search Based Heuristic for the Unrestricted Block Relocation Problem}

\author{Dominique Feillet\footnote{Ecole des Mines de Saint-Etienne and LIMOS UMR CNRS 6158, CMP Georges Charpak, Gardanne, F-13541 France, feillet@emse.fr},
   Sophie N. Parragh\footnote{Institute of Production and Logistics Management, Johannes Kepler University Linz, Austria, sophie.parragh@jku.at}, Fabien Tricoire\footnote{Institute of Production and Logistics Management, Johannes Kepler University Linz, Austria, fabien.tricoire@jku.at}}

\date{}

\maketitle

\begin{abstract}
The unrestricted block relocation problem is an important optimization
problem encountered at terminals, where containers are stored in stacks.  It consists in determining the minimum number of container moves so as to empty the considered bay following a certain retrieval sequence. A container move can be either the retrieval of a container or the relocation of a certain container on top of a stack to another stack. The latter types of moves are necessary so as to provide access to containers which are currently not on top of a stack. They might also be useful to prepare future removals. In this paper, we propose the first local search type improvement heuristic for the block relocation problem. It relies on a clever definition of the state space which is explored by means of a dynamic programming algorithm so as to identify the locally optimal sequence of moves of a given container. Our results on large benchmark instances reveal unexpectedly high improvement potentials (up to 50\%) compared to results obtained by state-of-the-art constructive heuristics.
\end{abstract}

\section{Introduction} \label{s:intro}

Container terminals are hubs connecting waterborne, rail, and road transportation networks. A more pronounced modal shift towards more sustainable means of transport, such as from road to rail or water, will put even more emphasis on their efficient operations. 
Within container terminals, containers awaiting further movement are
usually stored in stacks. This is due to restricted amounts of space,
as well as to restricted operating areas of handling equipment, such
as gantry cranes. Very often, the sequence in which containers arrive is
completely different from the sequence in which they need to be
retrieved; moreover, it is not possible to take into account the
sequence in which they will be retrieved while they arrive. This entails that in order to retrieve a container which does not end up on top of a stack, other containers have to be moved (to other stacks) in the same bay. Assuming the sequence in which the stacked containers need to be retrieved to be known, the aim is to minimize the number of unproductive movements
(i.e., the number of reshuffles). The underlying optimization problem is known as the container relocation problem (CRP) \citep{forster2012tree} or the block(s) relocation problem (BRP) \citep{caserta2011applying} in the literature. In its basic version, first introduced by \citet{kim2006heuristic}, unproductive movements may only concern those containers which are on top of the container which needs to be retrieved next. It is often referred to as the restricted container (or block) relocation problem (R-BRP) \citep[see, e.g.,][]{galle2018new}. In this paper we study the BRP without such a restriction. This means that, in anticipation of the retrieval sequence, unproductive relocations may concern containers on top of any of the stacks. In the literature, this version is usually referred to as the unrestricted BRP (U-BRP) and several authors have proposed heuristic as well as exact solution methods to solve the U-BRP. In this paper, we propose the first local search based improvement method for the U-BRP.

The BRP has received considerable attention in the literature. For a literature review and a classification scheme considering all kinds of problems arising in situations where containers are organized in stacks, we refer to \citet{lehnfeld2014loading}. This review covers the literature until 2014. Therefore, we will focus here on those works which have appeared later than that. We will first give a brief overview of contributions which study the R-BRP and we will then discuss recent works on the unrestricted version in further detail.

Among the more recent works on the restricted BRP is the one of \citet{jovanovic2014chain}. They propose a chain heuristic which introduces a look-ahead feature and is thus able to improve on the results of basic greedy algorithms developed previously. More recently, \citet{exposito2015exact} have proposed a mathematical model for the R-BRP and a branch-and-bound algorithm which can be truncated and thus used as a heuristic. 
\citet{eskandari2015notes} propose corrections and improvements to one of the formulations proposed in \citet{caserta2012mathematical} and present improved results for the R-BRP and 
\citet{ku2016abstraction} develop an exact method which relies on
clever ways to reduce the search space as well as on a bi-directional search scheme. Very recently, \citet{galle2018new} have propose a new mathematical formulation relying on a binary encoding of different configurations.

For the unrestricted BRP studied in this paper, \citet{jin2015solving} propose a look-ahead heuristic, in which a tree search algorithm is called so as to anticipate the impact of the next relocation. They are able to improve on the results of \citet{forster2012tree} and \citet{zhu2012iterative}. 

\citet{tricoire2018new} design a branch-and-bound algorithm and several heuristics for the U-BRP, relying on the notion of \emph{safe moves} and \emph{decreasing sequences}, and incorporate them into metaheuristic search schemes which they call \emph{rake search} and \emph{pilot method}. The branch-and-bound algorithm relies on a new lower bound, improving on the one of \citet{forster2012tree}, and it employs the loop idea of \citet{tanaka2015dominance}, in which the lower bound is iteratively increased by one.

Very recently, \citet{tanaka2018exact} have designed a
branch-and-bound algorithm for the unrestricted BRP. It relies on new
dominance properties concerning partial sequences of relocations which
do not appear in optimal solutions, as well as an improved lower
bound. Benchmark instances with at most 10 stacks and 50 items are
consistently solved to optimality with stacking height restrictions within 30 minutes of computation
time.

Also \citet{azari2017decreasing} rely on a branch-and-bound based heuristic algorithm. In contrast to other studies, they aim at minimizing the total working time of the crane.

New lower bounds for the U-BRP have been proposed and used in an A* type algorithm by \citet{quispe2018brp}. The first new bound is a combinatorial lower bound, the second relies on pattern databases.

The pre-marshalling problem concerns the re-ordering of containers
stored in stacks such that no reshuffling is necessary during
retrieval. New unifying problem formulations of the pre-marshalling problem, the R-BRP and the U-BRP have recently been developed by \citet{DEMELODASILVA201840}. While the new formulations are more efficient than previous mathematical programming based approaches the authors acknowledge that tailored branch-and-bound algorithms still provide superior performance in terms of run time as well as number of instances solved to optimality.

The pre-marshalling problem is also subject to investigation in \citet{jovanovic2017multi}. The authors propose a multi-heuristic method, which includes a correction step. This correction step detects consecutive relocations of the same container and merges them. The idea underlying this step is similar to what motivated us to develop a local search algorithm.

\citet{prandtstetter2013dynamic} developed a dynamic programming based branch-and-bound algorithm for the pre-marshalling problem. To reduce the number of states, a heuristic rule for determining equivalent states is introduced and used to obtain a heuristic algorithm.

Very recently \citet{tanaka2018prp} have proposed a new branch-and-bound algorithm for the pre-marshalling problem solving a number of previously unsolved instances to optimality.  \citet{perreno2019prp} develop new integer programming models as well as an iterative solution method which, in each iteration, increases the lower bound on the maximum number of moves needed to rearrange the bays by one until a feasible and thus optimal solution has been found.

Generalizations of the BRP involve, e.g., the work of \citet{hakan2014mathematical}, who introduce the dynamic container relocation problem in which containers are received and retrieved from a single yard-bay. They propose three different heuristics. \citet{ku2016container} propose to incorporate departure time windows and they develop a stochastic dynamic programming model, since the retrieval sequence of containers belonging to the same departure time window is unknown.

Summarizing the above, it becomes clear that, currently, three streams
of research exist. The first stream focuses on the development of
efficient mathematical models, the second stream concerns the design
of exact methods, mostly based on the branch-and-bound concept, while
the third stream of research focuses on the identification of
heuristic rules to obtain good but not necessarily optimal solutions
quickly. In this paper we open up a fourth stream of research: we
propose a local search based heuristic algorithm for the U-BRP, relying on dynamic programming, which allows to improve any heuristically generated solution. It locally reoptimizes the sequence of unproductive moves (or relocations) of a given container, with the objective of reducing their number. Relocations are not modified for other containers.

The remainder of the paper is organized as follows. Section~\ref{s:problem} gives a more detailed problem definition. In Section~\ref{s:LS} we present our local search based improvement heuristic. Numerical experiments are detailed and analyzed in Section \ref{s:exp}. Section \ref{s:conclusion} concludes the paper and draws some possible perspectives.

\section{Problem definition}\label{s:problem}

We now provide a definition of the U-BRP. Given a set of $N$ blocks or
containers which are stored in $W$ stacks of maximum height $H_{max}$
in a bay, the aim is to retrieve the containers according to a certain
sequence. This sequence is represented by consecutive numbers from $1$
to $N$ assigned to the containers, where container number 1 has to be
retrieved first and container number $N$ last.
A container can only be retrieved if it sits on top of a stack; if
there are other containers sitting above it, they must be moved to
other stacks first.
Moving a container to another stack is what we call a \emph{relocation}. The aim is to minimize the total number of relocations $R$ necessary to retrieve all containers. The total number of container movements is then given by $N+R$ (retrieval plus relocation movements). We call {\it configuration} the layout of the bay at a given stage of the removal process. Each move (relocation or removal of a container) induces a new configuration. These configurations are numbered from $1$ to $N+R+1$, where configuration $1$ represents the initial layout of the bay and configuration $N+R+1$ the final (empty) layout. Note that this final configuration is numbered $N+R+1$ because it results from $N$ retrievals and $R$ relocations. A solution $\mathcal S$ to the U-BRP represents the sequence of $N+R+1$ configurations.

\section{A local search algorithm for the U-BRP}\label{s:LS}

As explained in Section \ref{s:intro}, our new local search operator
$OPT(n)$ reoptimizes the relocations of container $n$ without
modifying the moves of any other container ($1 \leq n \leq N$). The locally best sequence  of moves for container $n$ is computed by dynamic programming. Given a starting solution $\mathcal S$,  we repeatedly apply $OPT(n)$ for all containers, until a local optimum is reached. The local search algorithm (LS) is detailed in Algorithm \ref{a:algoLS}. 

In order to avoid useless calls to $OPT(n)$, a lower bound $LB_n$ is
precomputed for each container. It gives the minimal number of moves
required for the container. This number is 1 if this
container blocks another container with higher priority, 0 otherwise. A container is skipped when its number of relocations in $\mathcal S$ (denoted $f_n$) is equal to $LB_n$ (Line \ref{a:testDP}).

\begin{algorithm}[ht]
	\begin{algorithmic}[1] 
		\STATE $improvement \leftarrow true$
		\WHILE{$improvement = true$}
		\STATE $improvement \leftarrow false$
		\FORALL{container $n \in \{1,\dots,N\}$}
		\IF{$f_n > LB_n$} \label{a:testDP}
		\STATE apply $OPT(n)$
		\IF{$OPT(n)$ enables reducing the number of relocations for container $n$}
		\STATE update  $\mathcal S$
		\STATE 	$improvement \leftarrow true$	
		\ENDIF
		\ENDIF
		\ENDFOR	
		\ENDWHILE
	\end{algorithmic}
	\caption{Local search algorithm (LS)} \label{a:algoLS}
\end{algorithm}

In what follows, we first detail the state space defined by our new local search operator $OPT(n)$
(Section \ref{s:LS_state_space}). Thereafter, in Section \ref{s:LS_opt}, we precise how the best path in this state-space is computed.

\subsection{State-space}  \label{s:LS_state_space}

In this section, we introduce the state space on which dynamic programming is applied. 
Let $\mathcal S$ be a BRP solution on which we apply $OPT(n)$. Recall that $\mathcal S$ represents the $N+R+1$ bay configurations, each defined by the retrieval or relocation of one container, starting from the initial bay configuration and ending with an empty bay.

We call {\it step} the transition from a configuration to the next in solution $\mathcal S$. Steps are numbered from $1$ to $N+R$. Step $k$ allows to pass from configuration $k$ to configuration $k+1$. 
We denote the operation constituting a transition or step $k$ by $\sigma^k = (\sigma_1^k, \sigma_2^k)$, where $\sigma^k_1$ denotes the original stack of the container which is moved  (in configuration $k$) and   $\sigma^k_2$ its new stack (in configuration $k+1$), $\sigma^k_2=``-''$ in case of a retrieval.

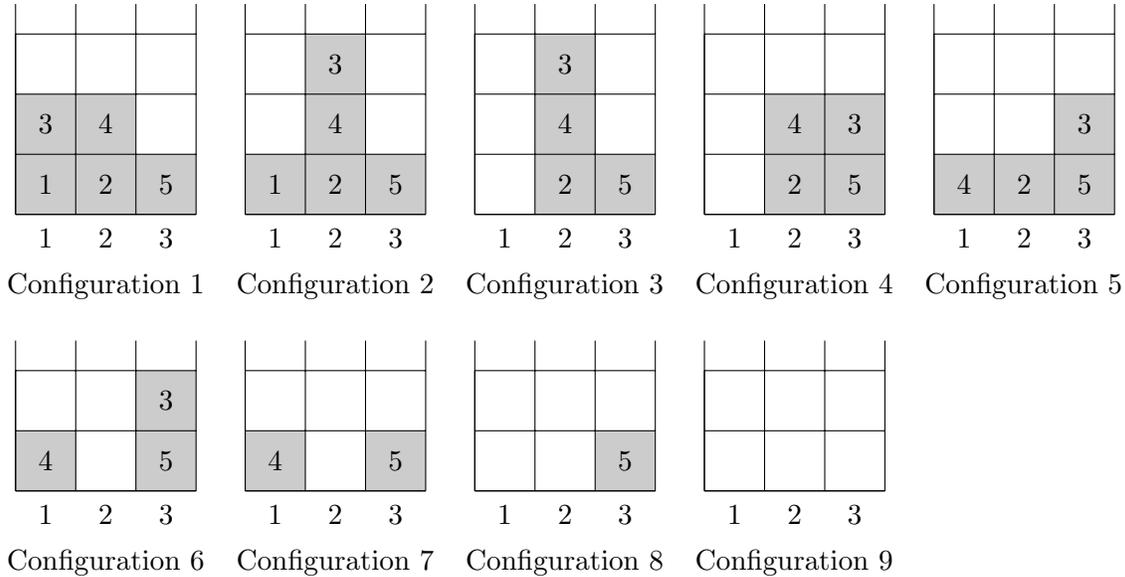
\begin{figure}[htbp] 
	\begin{center}
\begin{tabular}{ccccc}	
\begin{tikzpicture}[scale=\myScale]
\drawThreeStacks{2}{2}{1}{3}{0}

\draw (0.5,0.5) node {$1$};
\draw (0.5,1.5) node {$3$};

\draw (1.5,0.5) node {$2$};
\draw (1.5,1.5) node {$4$};

\draw (2.5,0.5) node {$5$};

\end{tikzpicture}
&
\begin{tikzpicture}[scale=\myScale]
\drawThreeStacks{1}{3}{1}{3}{0}

\draw (0.5,0.5) node {$1$};

\draw (1.5,0.5) node {$2$};
\draw (1.5,1.5) node {$4$};
\draw (1.5,2.5) node {$3$};

\draw (2.5,0.5) node {$5$};

\end{tikzpicture}
&
\begin{tikzpicture}[scale=\myScale]
\drawThreeStacks{0}{3}{1}{3}{0}

\draw (1.5,0.5) node {$2$};
\draw (1.5,1.5) node {$4$};
\draw (1.5,2.5) node {$3$};

\draw (2.5,0.5) node {$5$};

\end{tikzpicture}
&
\begin{tikzpicture}[scale=\myScale]
\drawThreeStacks{0}{2}{2}{3}{0}

\draw (1.5,0.5) node {$2$};
\draw (1.5,1.5) node {$4$};

\draw (2.5,0.5) node {$5$};
\draw (2.5,1.5) node {$3$};

\end{tikzpicture}
&
\begin{tikzpicture}[scale=\myScale]
\drawThreeStacks{1}{1}{2}{3}{0}

\draw (0.5,0.5) node {$4$};

\draw (1.5,0.5) node {$2$};

\draw (2.5,0.5) node {$5$};
\draw (2.5,1.5) node {$3$};

\end{tikzpicture}\\
Configuration 1 & Configuration 2 & Configuration 3 & Configuration 4 & Configuration 5\\
~\\

\begin{tikzpicture}[scale=\myScale]
\drawThreeStacks{1}{0}{2}{2}{0}

\draw (0.5,0.5) node {$4$};

\draw (2.5,0.5) node {$5$};
\draw (2.5,1.5) node {$3$};

\end{tikzpicture}
&
\begin{tikzpicture}[scale=\myScale]
\drawThreeStacks{1}{0}{1}{2}{0}

\draw (0.5,0.5) node {$4$};

\draw (2.5,0.5) node {$5$};

\end{tikzpicture}
&
\begin{tikzpicture}[scale=\myScale]
\drawThreeStacks{0}{0}{1}{2}{0}

\draw (2.5,0.5) node {$5$};

\end{tikzpicture}
&
\begin{tikzpicture}[scale=\myScale]
\drawThreeStacks{0}{0}{0}{2}{0}

\end{tikzpicture}\\
Configuration 6 & Configuration 7 & Configuration 8 & Configuration 9\\

\end{tabular}
\end{center}
	\caption{Configurations of a given solution $\mathcal S$ for a
          BRP instance with $W = 3$ and $N = 5$.} \label{f:example}
\end{figure}

\begin{table}[htbp]
	\begin{center}
		\begin{tabular}{ccccccc} \cline{1-3} \cline{5-7}  
			\multicolumn{2}{c}{Step} & $(\sigma_1,\sigma_2)$ & \hspace{1cm} & \multicolumn{2}{c}{Step} & $(\sigma_1,\sigma_2)$   \\  \cline{1-3} \cline{5-7}  
			1 & Relocate 3 & $(1,2)$ &  & 5 & Retrieve 2 & $(2,-)$ \\
			2 & Retrieve 1 & $(1,-)$ & & 6 & Retrieve 3 & $(3,-)$\\
			3 & Relocate 3 & $(2,3)$ & & 7 & Retrieve 4 & $(1,-)$\\
			4 & Relocate 4 & $(2,1)$ & & 8 & Retrieve 5 & $(3,-)$\\  \cline{1-3} \cline{5-7}  
		\end{tabular}				
	\end{center}
	\caption{Steps for the solution $\mathcal S$ of Figure \ref{f:example}} \label{t:example}
\end{table}

Figure \ref{f:example} and Table \ref{t:example} illustrate the concepts of configuration and steps for a given solution of a small BRP instance.

\subsubsection{Removal of container $n$ from solution $\mathcal S$}

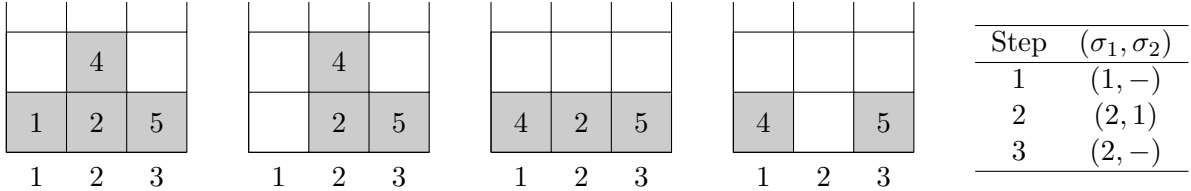
\begin{figure}[htbp] 
	\begin{center}
	\begin{tabular}{ccccc}
		\begin{minipage}{2.8cm}
		\begin{tikzpicture}[scale=\myScale]
		\drawThreeStacks{1}{2}{1}{2}{0}
		
		\draw (0.5,0.5) node {$1$};
		
		\draw (1.5,0.5) node {$2$};
		\draw (1.5,1.5) node {$4$};
		
		\draw (2.5,0.5) node {$5$};
		
		\end{tikzpicture}
		\end{minipage}
		&
		\begin{minipage}{2.8cm}		\begin{tikzpicture}[scale=\myScale]
		\drawThreeStacks{0}{2}{1}{2}{0}
		
		\draw (1.5,0.5) node {$2$};
		\draw (1.5,1.5) node {$4$};
		
		\draw (2.5,0.5) node {$5$};
		
		\end{tikzpicture}
		\end{minipage}
		&
		\begin{minipage}{2.8cm}		\begin{tikzpicture}[scale=\myScale]
		\drawThreeStacks{1}{1}{1}{2}{0}

		\draw (0.5,0.5) node {$4$};
		
		\draw (1.5,0.5) node {$2$};
		
		\draw (2.5,0.5) node {$5$};
		
		\end{tikzpicture}
		\end{minipage}
		&
		\begin{minipage}{2.8cm}
		\begin{tikzpicture}[scale=\myScale]
		\drawThreeStacks{1}{0}{1}{2}{0}
		
		\draw (0.5,0.5) node {$4$};
		
		\draw (2.5,0.5) node {$5$};
		
		\end{tikzpicture}
		\end{minipage}
		& 
		\begin{minipage}{2.8cm}
			\begin{tabular}{cc}
				\hline
				Step & $(\sigma_1,\sigma_2)$\\ \hline
				1 & $(1,-)$ \\
				2 & $(2,1)$ \\
				3 & $(2,-)$ \\ \hline
			\end{tabular}
		\end{minipage}
	\end{tabular}
\end{center}
	\caption{$\mathcal{S}^{-3}$  for solution $ \mathcal S$ of Figure \ref{f:example}} \label{f:Sminusn}
\end{figure}

We introduce a sequence of configurations and steps, that we call $\mathcal{S}^{-n}$, and whose definition is detailed subsequently. Intuitively,  $\mathcal{S}^{-n}$ only considers the part of solution $\mathcal S$ that occurs before the retrieval of container $n$ and modifies this sub-solution by removing container $n$ from the bay. More precisely,  $\mathcal{S}^{-n}$ is obtained by the following constructive algorithm:
\begin{enumerate}
	\item $\mathcal{S}^{-n}$ is initialized with no configuration;
	\item Configurations and steps of solution $\mathcal{S}$ are
          appended to $\mathcal{S}^{-n}$:
	\begin{itemize}
		\item starting from configuration $1$ and going up to the last configuration before the retrieval of container $n$,
		\item ignoring configurations that result from the relocation of container $n$.
		\end{itemize}
	\item Container $n$ is then removed from all these configurations, {\it i.e.}, for each configuration:
	\begin{itemize}
		\item the position where $n$ is located is first emptied,
		\item the tier of every container located above $n$ is decreased by 1.
	\end{itemize} 
\end{enumerate} 
Steps are defined accordingly. Figure \ref{f:Sminusn} depicts the configurations and steps of $\mathcal{S}^{-3}$  for the example of Figure \ref{f:example}. Configurations 1 to 6 of Figure \ref{f:example} are considered, in this order. Configurations 2 and 4 are skipped because they follow the relocation of container 3. In the four remaining configurations, container 3 is removed.

We denote by $M$ the number of configurations in $\mathcal{S}^{-n}$. These configurations are numbered from 1 to $M$, and the steps between these configurations from 1 to $M-1$  ($M=4$ in the example of Figure \ref{f:Sminusn}).

We denote by $h(s,t)$ the height of stack $s$ in configuration $t$ of  $\mathcal{S}^{-n}$. For example, in the configurations depicted in Figure \ref{f:Sminusn}, $h(2,1)=h(2,2)=2$, $h(2,3)=1$ and  $h(2,4)=0$.

\subsubsection{Definition of the state space graph}

The state space graph is defined by states (nodes) and transitions (arcs) between these states.

\paragraph{States (nodes)} of the state-space are named $STATE(t,s,h)$, where $t \in \{1,\dots,M\}$ indicates a configuration in $\mathcal S^{-n}$, $s \in \{1,\dots,W\}$ indicates a stack in the bay and $h \in \{1,\dots,H_{max}\}$ a tier in this stack. 

\begin{figure}[htbp] 
	\begin{center}
\begin{tabular}{cccc}	
\begin{tikzpicture}[scale=\myScale]
\drawThreeStacks{0}{2}{1}{3}{0}

\draw (1.5,0.5) node {$2$};
\draw (1.5,1.5) node {$4$};

\draw (2.5,0.5) node {$5$};

\end{tikzpicture}
&
\begin{tikzpicture}[scale=\myScale]
\drawThreeStacks{0}{2}{1}{3}{0}

\draw (0.5,1.5) node {\bf 3};

\draw (1.5,0.5) node {$2$};
\draw (1.5,1.5) node {$4$};

\draw (2.5,0.5) node {$5$};

\end{tikzpicture}
&
\begin{tikzpicture}[scale=\myScale]
\drawThreeStacks{0}{3}{1}{3}{0}

\draw (1.5,0.5) node {$2$};
\draw (1.5,1.5) node {\bf 3};
\draw (1.5,2.5) node {$4$};

\draw (2.5,0.5) node {$5$};

\end{tikzpicture}
&
\begin{tikzpicture}[scale=\myScale]
\drawThreeStacks{0}{2}{2}{3}{0}

\draw (1.5,0.5) node {$2$};
\draw (1.5,1.5) node {$4$};

\draw (2.5,0.5) node {$5$};
\draw (2.5,1.5) node {\bf 3};

\end{tikzpicture}
\\
Configuration 2 & $STATE(2,1,2)$ &  $STATE(2,2,2)$ & $STATE(2,3,2)$ \\
\end{tabular}
\end{center}
	\caption{Some examples of states for configurations of  $\mathcal{S}^{-3}$  depicted in  Figure \ref{f:Sminusn}} \label{f:state}
\end{figure}
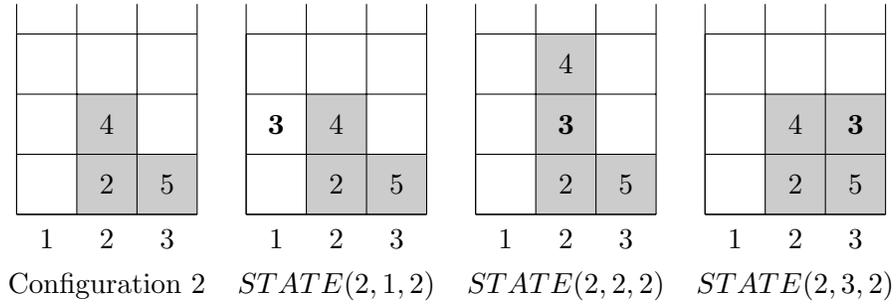

$STATE(t,s,h)$ represents a bay configuration based on configuration $t$ of solution $\mathcal{S}^{-n}$, modified by inserting container $n$ in stack $s$ and at tier $h$.  The container initially located at position $(s,h)$ in configuration $t$ and containers above have their tier augmented by $1$. Figure \ref{f:state} gives some examples of states.

A state $STATE(t,s,h)$ is said to be feasible if the following conditions hold. 
\begin{itemize}
	\item States with $2 \leq t \leq M-1$:
\begin{enumerate}
	\item Container $n$ is not floating above other containers: $h \leq h(s,t)+1$;
	\item A slot is available in the stack to insert container $n$: $h(s,t) < H_{max}$;
	\item Container $n$ is not on top of the container that is to be moved in the next step: if $\sigma^t_1 = s$, $h \leq h(s,t) $;
	\item Container $n$ does not complete the stack that receives a container at the next step: if $\sigma^t_2 = s$, $h(s,t) + 1 < H_{max}$.
\end{enumerate}
\item States with $t=1$: 
\begin{description}
\item The only feasible state is $STATE(1,s^0_n,h^0_n)$ that represents the initial layout of the bay, where $(s^0_n,h^0_n)$ are the initial coordinates of $n$. This state is called initial state.
\end{description}
\item States with $t=M$: 
\begin{enumerate}
	\item Container $n$ is exactly on top of a stack: $h = h(s,t)+1$;
	\item A slot is available in this stack: $h(s,t) < H_{max}$.
\end{enumerate}
Feasible states with $t = M$ are thus those for which container $n$ can be retrieved. These states are called final states.
\end{itemize}
For example, the last two states on Figure \ref{f:state} are feasible,
but the first one is not.

\paragraph{Transitions (arcs)} between states are defined as follows.
Given a feasible $STATE(t,s,h)$, transitions from this state are only defined to a subset of feasible states of the next configuration, that is, states of the form $STATE(t+1,s',h')$. They consist of relocating (or not) container $n$ and applying step $t$ of solution $\mathcal{S}^{-n}$.

Let us consider a feasible state $STATE(t,s,h)$ with $t<M$. Transitions from this state are defined as follows:
\begin{itemize}
	\item A transition of cost $0$ to $STATE(t+1,s,h)$ is added if $STATE(t+1,s,h)$ is feasible.
	\item If $h=h(s,t)+1$, a transition of cost $1$ to $STATE(t+1,s',h(s',t+1)+1)$ is added for all stacks $s' \in \{1,\dots,S\}\setminus\{s\}$ if $STATE(t+1,s',h(s',t+1)+1)$ is feasible.
\end{itemize}
The first case corresponds to a strategy where container $n$ is not relocated and step $t$ is simply applied. In the second case, container $n$ is relocated to stack $s'$ before applying step $t$.

The optimal set of relocations for container $n$ is then obtained by
computing a shortest path in the graph defined by this state space, from the initial state $STATE(1,s^0_0,h^0_n)$ to any of the final states. 

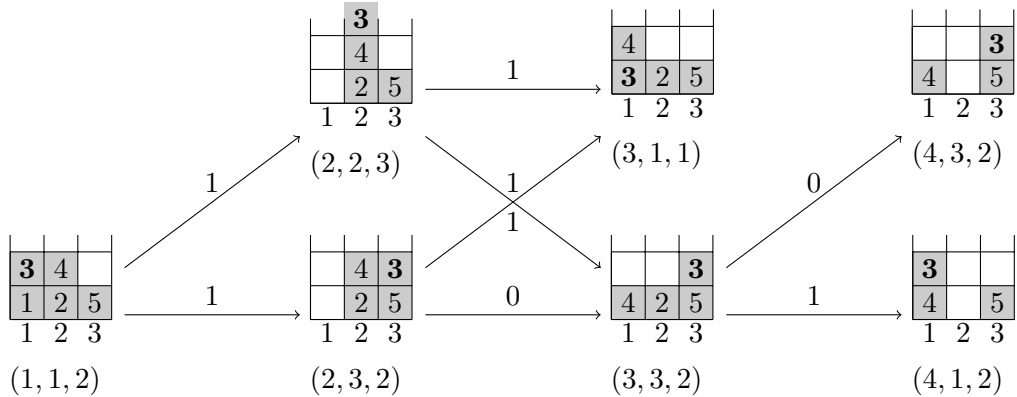
\begin{figure}[htbp] 
	\begin{center}
\begin{tikzpicture}[scale=0.5]

\node(012) at (0,0) {
		\begin{minipage}{1.4cm}
		\begin{tikzpicture}[scale=0.45]
		\drawThreeStacks{2}{2}{1}{2}{0}
		
		\draw (0.5,0.5) node {$1$};
		
		\draw (1.5,0.5) node {$2$};
		\draw (1.5,1.5) node {$4$};
		
		\draw (2.5,0.5) node {$5$};
		\draw (0.5,1.5) node {\bf3};
		\end{tikzpicture}
		$(1,1,2)$
		\end{minipage}};
\node(123) at (8,6) {
	\begin{minipage}{1.4cm}		
	\begin{tikzpicture}[scale=0.45]
	\drawThreeStacks{0}{3}{1}{2}{0}
		
	\draw (1.5,0.5) node {$2$};
	\draw (1.5,1.5) node {$4$};
		
	\draw (2.5,0.5) node {$5$};
	\draw (1.5,2.5) node {\bf3};
	\end{tikzpicture}
	$(2,2,3)$
	\end{minipage}};
\node(132) at (8,0) {
		\begin{minipage}{1.4cm}		
		\begin{tikzpicture}[scale=0.45]
		\drawThreeStacks{0}{2}{2}{2}{0}
		
		\draw (1.5,0.5) node {$2$};
		\draw (1.5,1.5) node {$4$};
		
		\draw (2.5,0.5) node {$5$};
		\draw (2.5,1.5) node {\bf3};
		\end{tikzpicture}
		$(2,3,2)$
		\end{minipage}
		};
\node(211) at (16,6) {
		\begin{minipage}{1.4cm}		
		\begin{tikzpicture}[scale=0.45]
		\drawThreeStacks{2}{1}{1}{2}{0}

		\draw (0.5,1.5) node {$4$};
		
		\draw (1.5,0.5) node {$2$};
		
		\draw (2.5,0.5) node {$5$};
		\draw (0.5,0.5) node {\bf3};
		\end{tikzpicture}
		$(3,1,1)$
		\end{minipage}
	};
\node(232) at (16,0) {
		\begin{minipage}{1.4cm}		
		\begin{tikzpicture}[scale=0.45]
		\drawThreeStacks{1}{1}{2}{2}{0}

		\draw (0.5,0.5) node {$4$};
		
		\draw (1.5,0.5) node {$2$};
		
		\draw (2.5,0.5) node {$5$};
		\draw (2.5,1.5) node {\bf3};
		\end{tikzpicture}
		$(3,3,2)$
		\end{minipage}
		};
\node(332) at (24,6) {
		\begin{minipage}{1.4cm}
		\begin{tikzpicture}[scale=0.45]
		\drawThreeStacks{1}{0}{2}{2}{0}
		
		\draw (0.5,0.5) node {$4$};
		
		\draw (2.5,0.5) node {$5$};
		\draw (2.5,1.5) node {\bf3};
		\end{tikzpicture}
		$(4,3,2)$
		\end{minipage}
		};
\node(312) at (24,0) {
		\begin{minipage}{1.4cm}
		\begin{tikzpicture}[scale=0.45]
		\drawThreeStacks{2}{0}{1}{2}{0}
		
		\draw (0.5,0.5) node {$4$};
		
		\draw (2.5,0.5) node {$5$};
		\draw (0.5,1.5) node {\bf3};
		\end{tikzpicture}
		$(4,1,2)$
		\end{minipage}
		};
\path[->] (012) edge node[above]{1} (123);
\path[->] (012) edge node[above]{1} (132);
\path[->] (123) edge node[above]{1} (211);
\path[->] (123) edge node[above]{1} (232);
\path[->] (132) edge node[below]{1} (211);
\path[->] (132) edge node[above]{0} (232);
\path[->] (232) edge node[above]{0} (332);
\path[->] (232) edge node[above]{1} (312);

\end{tikzpicture}
\end{center}
	\caption{The state space graph for optimizing the relocations of container 3 of solution $\mathcal S$ given in Figure \ref{f:example}} \label{f:statespace}
\end{figure}

The state-space graph for our example is given in Figure \ref{f:statespace}. 
Obviously, from STATE(3,1,1) no transition to a feasible final state ($t=M$, $M = 4$) is possible. 

\subsection{Local search operator $OPT(n)$: exploration}  \label{s:LS_opt}

In the following, we explain the algorithm to explore the defined state-space so as to identify the best sequence of relocations for a given item. Thereafter, we discuss the implemented speedups.

\subsubsection{Algorithm}

To find the min-cost path in the state-space graph, this graph is not explicitly constructed. Instead, we perform a breadth-first search from the initial state. The algorithm is detailed in Algorithm \ref{a:algoDP} and explained hereafter.

\begin{algorithm}[ht]
	\begin{algorithmic}[1] 
		\STATE  $f(1,s^n_0,h^n_0) \leftarrow 0$
		\STATE $\mathcal Q \leftarrow \{ (s^n_0,h^n_0)\}$
		\FOR{t=1 \TO M-1}
			\STATE $\mathcal Q' \leftarrow \emptyset$
			\WHILE{$\mathcal Q \neq \emptyset$} \label{a:algoDP_while}
				\STATE $(s,h) \leftarrow \mbox{pop}(\mathcal Q)$
				\FORALL{$(s',h')$ such that  $STATE(t+1,s',h')$ is a successor of $STATE(t,s,h)$}  \label{a:algoDP_for}
					\STATE update $f(t+1, s',h')$ \label{a:algoDP_update}
					\STATE $\mathcal Q' \leftarrow \mathcal Q' \cup \{(s',h')\}$ \label{a:algoDP_add}
				\ENDFOR \label{a:algoDP_endfor}
			\ENDWHILE \label{a:algoDP_endwhile}
			\STATE $\mathcal Q \leftarrow \mathcal Q'$
		\ENDFOR	
	\end{algorithmic}
	\caption{Dynamic programming algorithm} \label{a:algoDP}
\end{algorithm}

Function $f$ represents the current minimal cost to reach a state. It is initialized to $0$ for the initial state.  A loop considers all the states that have been reached for a given value of $t$ and determines the reachable states for $t+1$. In order to do so,
two queues $\mathcal Q$ and $\mathcal Q'$ are used. $\mathcal Q$ contains the feasible states for configuration $t$. $\mathcal Q'$ is initially empty and progressively enriched with feasible successors of states in $\mathcal Q$. When $\mathcal Q'$ is complete,
i.e., all feasible successor states have been generated, $\mathcal Q'$ 
is copied to $\mathcal Q$ and the procedure is repeated for the next configuration. At line \ref{a:algoDP_update}, the value associated with a state is updated if a better path reaching this state has been found. 

In order to evaluate in constant time if a state is feasible, values $h(s,t)$ are precomputed for all stacks $s$ and configurations $t$. At Line \ref{a:algoDP_for}, all stacks $s'$ are tried for the generation/improvement of states; given $s'$, a single value $h'$ obtained from $h(s',t)$ is considered. The complexity of the loop (Lines \ref{a:algoDP_for}-\ref{a:algoDP_endfor}) is thus $W$ times the complexity of the update function. In order to achieve a complexity $O(1)$ for this function, two matrices of size $W \times H_{max}$ are introduced. The matrices store state information for configurations $t$ and $t+1$, respectively. With these matrices, $f(t+1, s',h')$ can be obtained in $O(1)$. The two matrices are reinitialized at the end of every iteration, when $t$ is increased. 

Queue $\mathcal Q$ has a maximal size equal to $W \times H_{max}$. All the elements in $\mathcal Q$ can {\it a priori} be transferred to $\mathcal Q'$ without an additional relocation, but only those elements with container $n$ on top of 
a stack allow a relocation.
The worst-case complexity of an iteration (Lines \ref{a:algoDP_while} to \ref{a:algoDP_endwhile}) is thus  $O(W \times H_{max} +W^2)$, {\it i.e.}, $O(W \times \max(H_{max},W))$. The worst-case complexity of Algorithm \ref{a:algoDP} is  finally $O(M \times W \times \max(H_{max},W))$.

\subsubsection{Speedups}

In order to reach a maximal efficiency for our local search operator,
we implemented several speedup techniques:
\begin{description}
	\item[Upper bound.] As the operator aims at decreasing the
          number of relocations for container $n$,  states that can
          only be reached with a large number of relocations of
          container $n$ are not interesting. Let $f_n$ be the number
          of relocations of $n$ in solution $\mathcal S$. If
          $f(t+1,s',h') \geq f_n$, $(s',h')$ is not added to $\mathcal
          Q'$ at Line \ref{a:algoDP_add}.
          Additionally, if $f(t+1,s',h') \geq f_n-1$ and state $(M,s',h')$ is not feasible, $(s',h')$ is also not added to $\mathcal Q'$. Indeed, under these conditions, one more relocation is needed for container $n$ before it can be retrieved.
	\item[Useless evaluations.] Applying a step $t$ of $\mathcal S^{-n}$ only modifies stacks $\sigma^t_1$ and $\sigma^t_2$.  Thus,  relocations of container $n$ for a given value of $t$ can be limited to the following two cases (except for $t=1$):
	\begin{enumerate}
		\item Container $n$ is located in stack $\sigma^{t-1}_1$. Then all relocations should be tried: they could not be performed at the previous steps.
		\item Container $n$ is located in another stack. Then, relocations for $n$ should be tried towards two stacks: $\sigma^{t-1}_1$ and $\sigma^{t-1}_2$. Other stacks have not been modified and acceptable relocations have already been generated. 
	\end{enumerate}
	\item[Aspiration.] Every time that the value of a state is updated (Line 	\ref{a:algoDP_update}), an attempt to generate  a final state is done. The principle is to identify if the state could be prolongated to a feasible final state without any relocation. Three conditions are evaluated:
	\begin{itemize} 
	\item $f(t+1,s',h') \leq f_n-1$,
	\item $h(s',t') < H_{max}$ for $t' \geq t+1$,
	\item $h(s',t') \geq h'-1$ for $t' \geq t+1$.
	\end{itemize}
The first condition ensures that the state would lead to an improvement. The second condition checks if there is room for container $n$ in stack $s'$ until it is retrieved. The third condition makes sure that containers below $n$ in stack $s'$  will not be moved until the retrieval (the height of the stack is always larger). If the three conditions hold, the dynamic programming algorithm is stopped. The sequence of relocations that leads to $(t+1,s',h')$ is returned. 
\end{description}

Note that the first two speedups do not modify the result of the
algorithm. The {\it aspiration} speedup on the contrary can modify the
sequence of relocations returned by the local search
procedure. Furthermore, this sequence is not necessarily optimal:
condition  $f(t+1,s',h') \leq f_n-1$ ensures that the sequence is
improving but some shorter paths might exist in the state-space.

Introducing the upper bound has no impact on the worst-case complexity of the algorithm.  With the {\it Useless evaluations} mechanism, the worst-case complexity of Lines \ref{a:algoDP_while}-\ref{a:algoDP_endwhile} falls to  $O(W \times H_{max} +W)$ and that of Algorithm \ref{a:algoDP} to $O(M \times W \times H_{max})$. The conditions for aspiration can be computed in constant time with a simple preprocessing phase computing values $\min_{t' \geq t} h(s,t)$ and $\max_{t' \geq t'} h(s,t')$ for all stacks $s$ and periods $t$. With these speedups, the new worst-case complexity of the algorithm is thus $O(M \times W \times H_{max})$.

\section{Numerical experiments}  \label{s:exp}

The proposed local search algorithm (LS) has been implemented in
C++. All tests have been carried out on Intel Xeon E5-2650 v2 CPUs
(25M Cache, 2.60 GHz). Each computing node has 2 processors with 8
cores each and 64 GB RAM. In total, up to 5 jobs run in parallel on
one node, sharing memory.

In order to analyze its performance, we 
apply our algorithm to three sets of instances. Two of which have been 
proposed by other authors (\citet{caserta2011applying} and \citet{tricoire2018new}) while the third set is new.
The small instances of
\citet{caserta2011applying} and the large instances proposed by
\citet{tricoire2018new}
follow the same scheme.
Let $H$ denote the maximum height of a stack
in an initial configuration, the size of the Caserta et al. instances ranges
from $(H,W)= (3,3)$ until $(H,W) = (10,10)$, the size of the Tricoire
et al. instances ranges from $(H,W)= (10,10)$ until $(H,W) =
(100,100)$, where we recall that $W$ is the number of
stacks. Furthermore, like in previous works, we consider two different
stack height limitations for each data set:  $H_{max} = H+2$ and
$H_{max} = unlimited$. Each instance class defined by $(H,W)$ contains
40 instances.

The purpose of the new set of wide instances is to mimic the actual height limitations of stacker cranes at container yards. Therefore, we set $H = 10$ and we only increase the width $W$. We start with $(H,W) = (10,10)$ and we increase it in steps of 10 until $(H,W) = (10,100)$. As previously, we consider two stack height limitations, namely $H_{max} = H+2$ and
$H_{max} = unlimited$, and we generate 40 instances for each instance class.

Since the aim of our LS is to improve on a previously generated
solution, we apply it to starting solutions produced by different
constructive heuristics. For the small Caserta instances, we use the
rake search and pilot method of \citet{tricoire2018new} as well as
two versions of the greedy look ahead heuristic (GLAH) of
\citet{jin2015solving}. 
For all other instances, we use the following four
fast construction heuristics:

\begin{itemize}
\item  JZW: evaluation subroutine of the greedy look ahead heuristic of \citet{jin2015solving}. This subroutine considers several different conditions for identifying the destination stack of a container which needs to be relocated, including the simultaneous relocation of another container if this seems advantageous.
\item LA-S-1: extended look-ahead heuristic of \citet{petering2013new}, where ``S-1'' defines the look ahead in terms of containers and here it is equal to $W-1$. It involves \emph{cleaning moves}. A cleaning move is a relocation of a container, whose relocation is not mandatory, to another stack from where it does not have to be relocated again.
\item SM-2: heuristic incorporating best safe 1-relocate moves and best safe 2-relocate moves \citep{tricoire2018new}. \emph{Safe relocations} are those which do not increase the proposed lower bound on the number of necessary relocations. 
\item SmSEQ-2: combination of decreasing sequence relocate (sequence
  length $\geq 1$) and SM-2  \citep{tricoire2018new}. \emph{Decreasing
    sequences} are sequences of containers which are stacked in the
  opposite order compared to the sequence in which they need to be
  retrieved. In certain cases, it is beneficial to relocate the whole
  sequence to some other stack $s$, even if that involves making
  room for it in $s$ first.
\end{itemize}

For all construction methods, we use the implementation
from~\cite{tricoire2018new}.

Many of the small instances from~\citet{caserta2011applying} are
solved optimally with constructive methods and the necessity of
developing complex local search operators is not obvious for
these instances. Even so, we evaluated our approach on these instances
and obtained only very limited improvements. These experiments are
reported in Tables~\ref{tab:advanced-caserta-unlimited}
and~\ref{tab:advanced-caserta-H+2} in the appendix. More
interesting results are obtained with larger instances like the ones
introduced in~\citet{tricoire2018new}.

\begin{figure}[htbp] 
	\centering \includegraphics[width=0.55\textwidth]{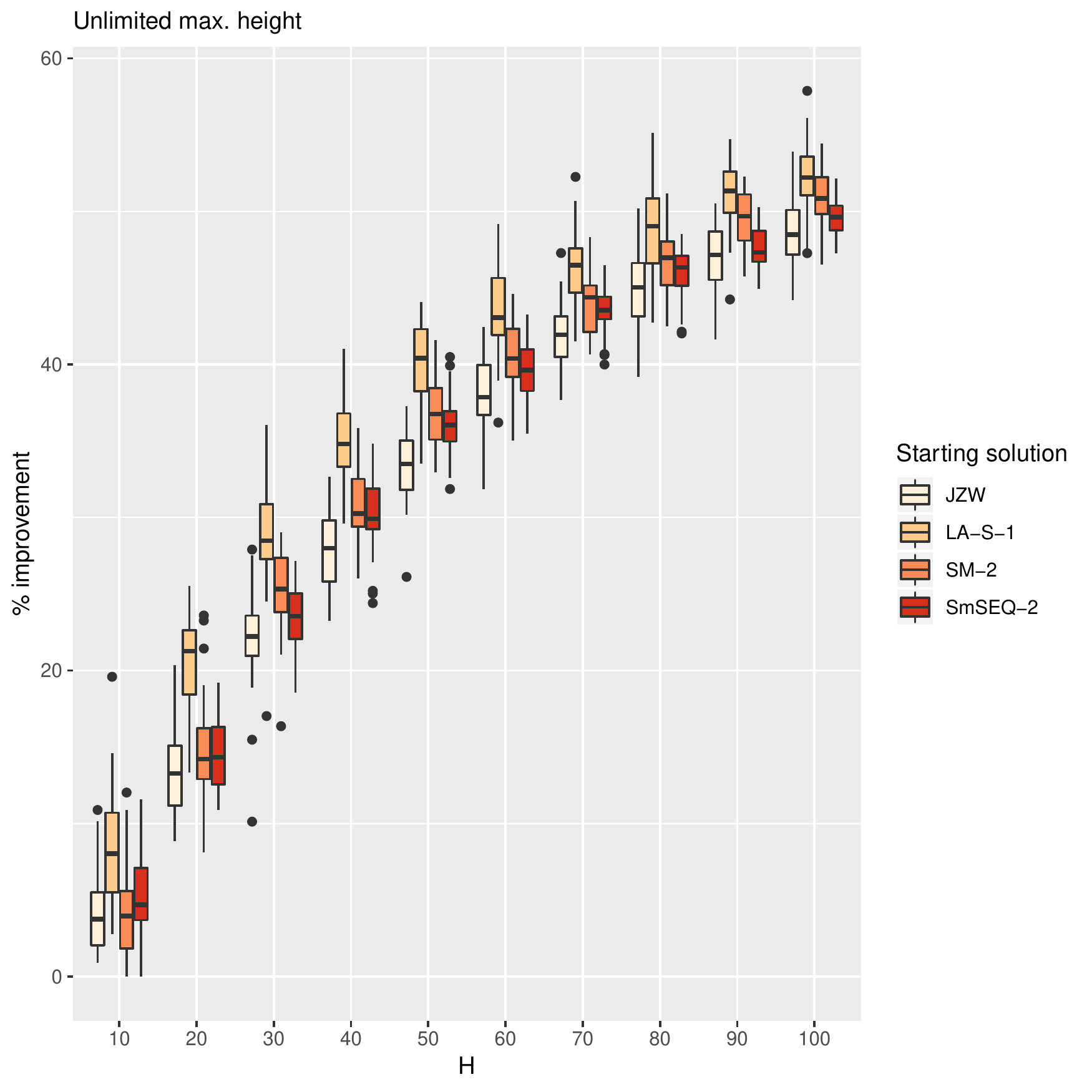}
	\caption{Average percentage improvement compared to starting solution for benchmark instances of \citet{tricoire2018new} with $H_{max}=unlimited$ and $H \in \{10,20,30,40,50,60,70,80,90,100\}$} \label{f:boxplot_percent_unlimited}
\end{figure}

\begin{figure}[htbp] 
	\centering \includegraphics[width=0.55\textwidth]{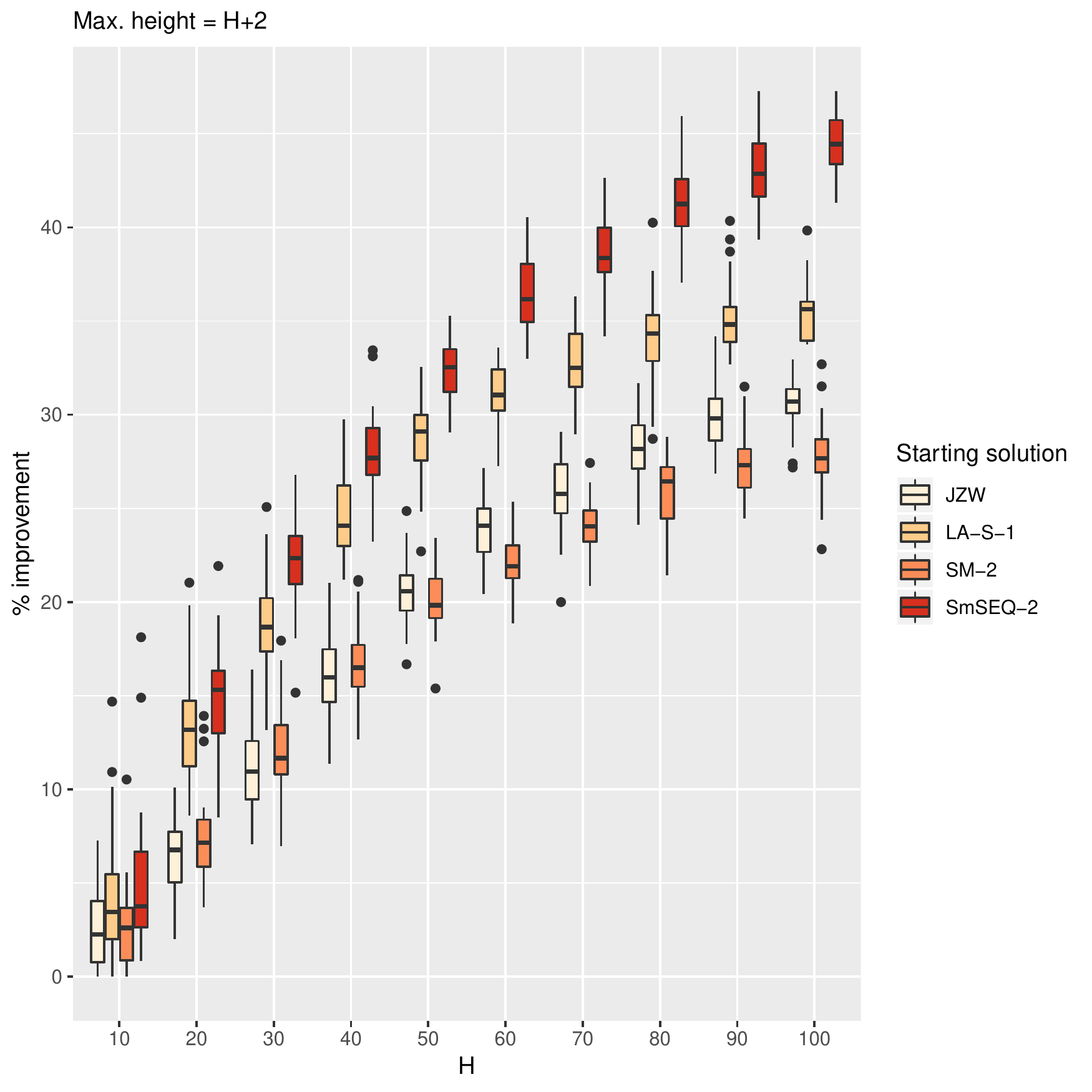}
	\caption{Average percentage improvement compared to starting solution for benchmark instances of \citet{tricoire2018new} with $H_{max}= H+2$ and $H \in \{10,20,30,40,50,60,70,80,90,100\}$} \label{f:boxplot_percent_H2}
\end{figure}

\begin{figure}[htbp] 
	\centering \includegraphics[width=0.55\textwidth]{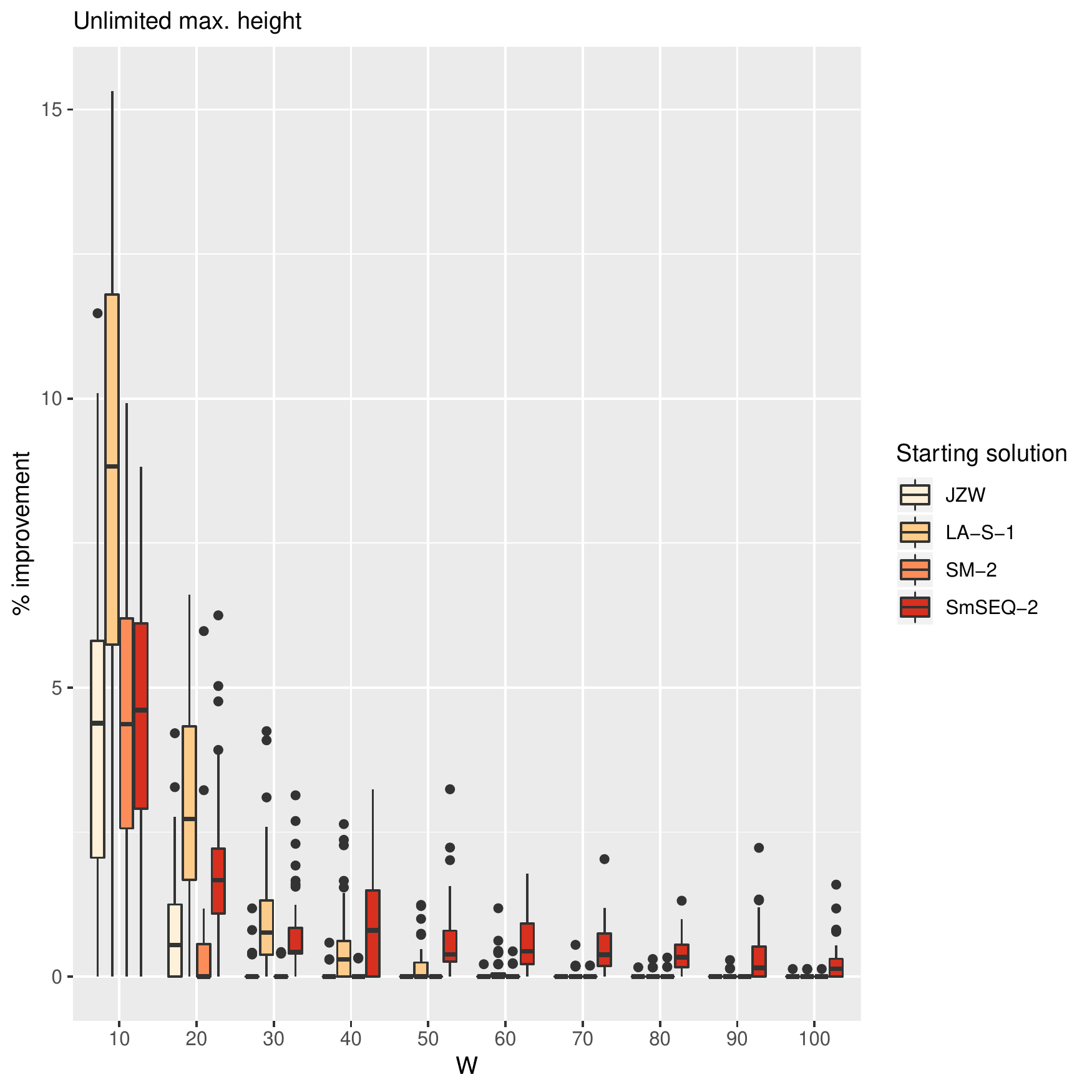}
	\caption{Average percentage improvement compared to starting solution for wide instances with $H_{max}=unlimited$ and $W \in \{10,20,30,40,50,60,70,80,90,100\}$} \label{f:boxplot_percent_unlimited-wide}
\end{figure}

\begin{figure}[htbp] 
	\centering \includegraphics[width=0.55\textwidth]{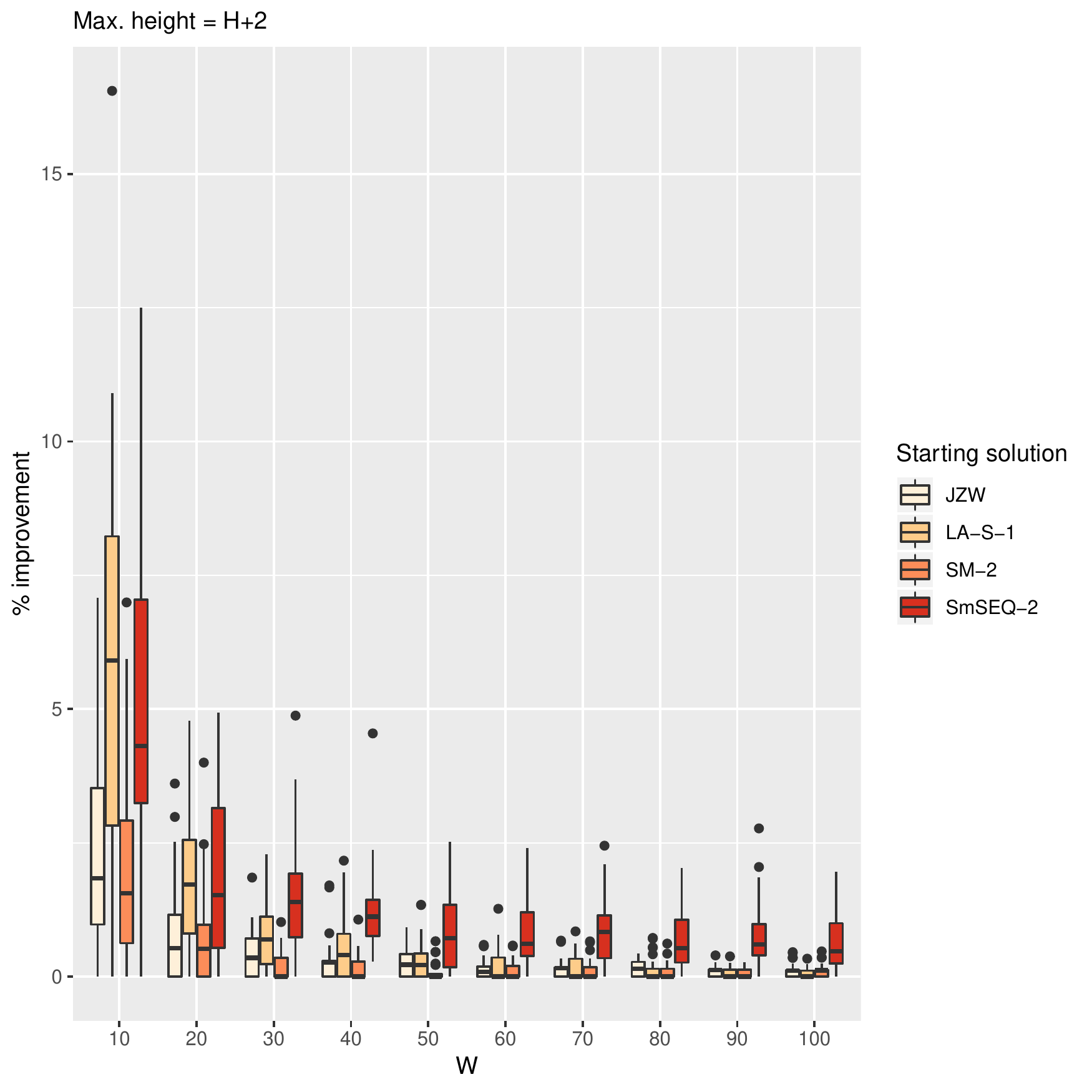}
	\caption{Average percentage improvement compared to starting solution for wide instances with $H_{max}= H+2$ and $W \in \{10,20,30,40,50,60,70,80,90,100\}$} \label{f:boxplot_percent_H2-wide}
\end{figure}

In a first stage, we analyze the improvement potential of our local
search algorithm. For this purpose, we plot the percentage improvement
(across 40 instances) per instance class using the respective objective
value obtained from the constructive heuristic as a base
line. Figures~\ref{f:boxplot_percent_unlimited}
--~\ref{f:boxplot_percent_H2-wide} provide the obtained results for
$H_{max} = unlimited$ and $H_{max} = H+2$, respectively.
For the instances of \citet{tricoire2018new} 
an obvious
trend can be observed: the larger the instance, the larger the
improvement by our local search algorithm. The improvements are consistently above $40\%$ for the largest instances without height limitation and above $20\%$ with $H_{max} = H+2$.
The respective average solution values per instance class, along with the average improvement compared to the starting solution, are given in Tables~\ref{tab:fast-large-avg-unlimited} and~\ref{tab:fast-large-avg-H+2} in the appendix.

For the wide instances the picture is different. We find improvements (detailed results can be found in Tables \ref{tab:fast-wide-avg-unlimited} and \ref{tab:fast-wide-avg-H+2} in the Appendix) but they are much less pronounced than for the large instances. Obviously, the larger the width of the instance the smaller the improvement potential.

\begin{figure}[htbp] 
  \centering \includegraphics[width=0.55\textwidth]{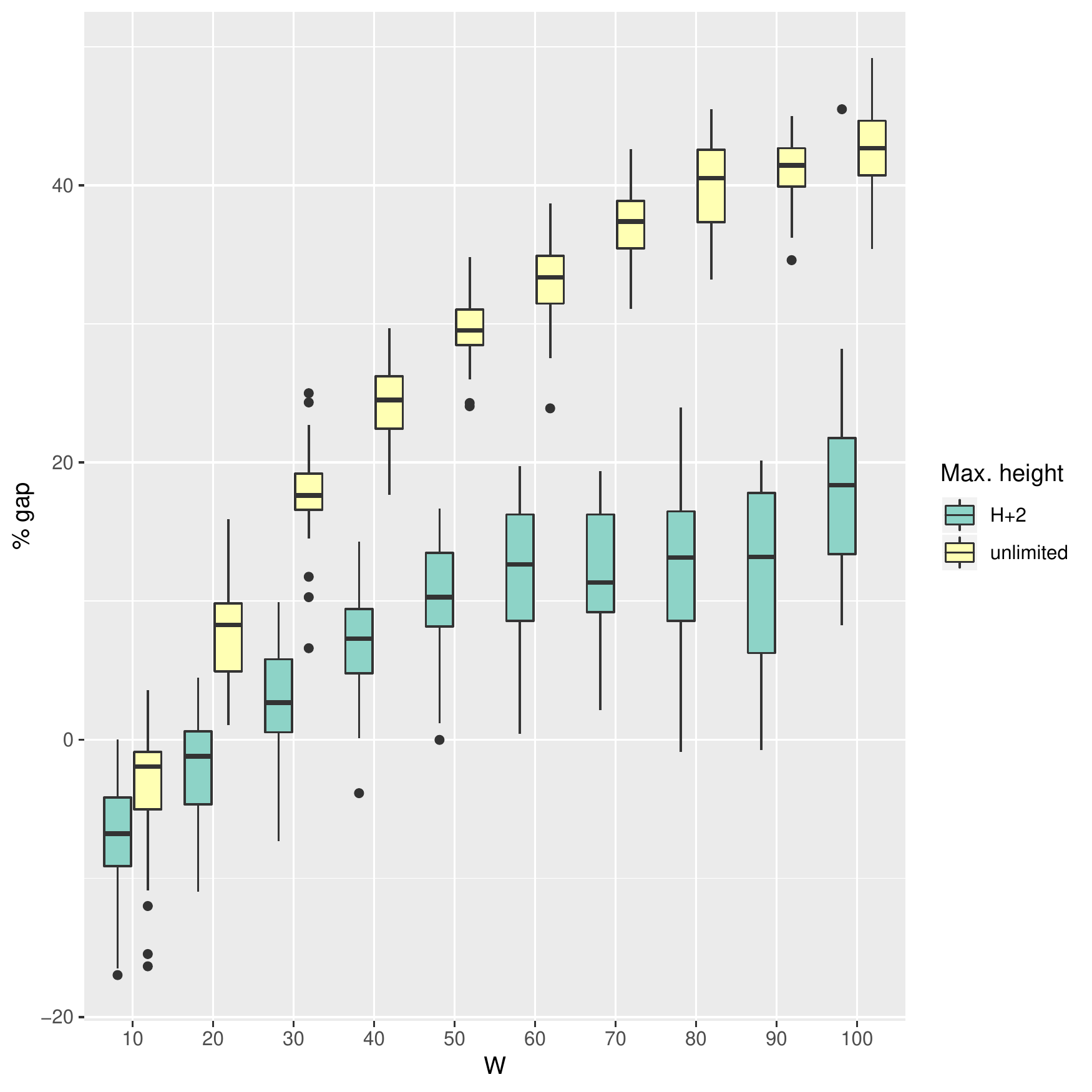}
  \caption{Worst percentage improvement compared to best constructive solution across all methods for benchmark instances of \citet{tricoire2018new}} \label{f:best_vs_worst}
\end{figure}

In a second stage, we consider for any given instance $x$ two values:
\begin{itemize}
\item $BB(x)$ is the best solution value for instance $x$ among all
  starting solutions, i.e. before LS
\item $WA(x)$ is the worst solution value for instance $x$ among all
  post-LS solutions, i.e. after LS.
\end{itemize}
In order to evaluate the overall performance of our
LS, we compare the $BB(x)$ against $WA(x)$ instance-wise. This
comparison (again in terms of percentage deviations) is given in
Figure~\ref{f:best_vs_worst}.
It illustrates that, in the case of the benchmark instances of
\citet{tricoire2018new} except for the
smallest instances, the worst solution obtained by LS is usually
better than the best solution identified with the best performing
constructive method. This improvement is more  pronounced for
instances without a height restriction, i.e. $H_{max} = unlimited$:
more than $30\%$ for instances with $H$ and $W$ both $\geq 70$. This
emphasizes the value of the proposed local search algorithm:
on these instances,
as long as LS is applied, it does not matter which
construction method is used to begin with, the end result will almost
always be better than what any construction method could do.

\begin{table}
\scriptsize
\centering
\begin{tabular}{cccccc}
\hline
$H$ & $W$ & \multicolumn{ 1 }{c}{ LA-S-1 } & \multicolumn{ 1 }{c}{ SM-2 } & \multicolumn{ 1 }{c}{ SmSEQ-2 } & \multicolumn{ 1 }{c}{ JZW } \\
\hline
10 & 10 &     0.00 &     0.00 &     0.00 &     0.00 \\
20 & 20 &     0.53 &     0.39 &     0.34 &     0.35 \\
30 & 30 &     7.73 &     6.14 &     3.97 &     5.01 \\
40 & 40 &    63.60 &    47.89 &    30.40 &    40.47 \\
50 & 50 &   323.99 &   254.74 &   162.90 &   219.89 \\
60 & 60 &  1210.38 &  1000.28 &   595.52 &   910.88 \\
70 & 70 &  3630.31 &  3101.77 &  1935.90 &  2836.54 \\
80 & 80 &  8398.51 &  7196.79 &  4524.30 &  6536.57 \\
90 & 90 & 18966.91 & 16862.56 & 10333.31 & 15286.99 \\
100 & 100 & 38592.69 & 34182.83 & 21072.38 & 31421.58 \\
\hline
\end{tabular}
\caption{CPU time in seconds of LS, using different methods for the starting solution. Tricoire instances, $H_{max} = unlimited$.}
\label{tab:fast-large-unlimited-CPU}
\end{table}
\begin{table}
\scriptsize
\centering
\begin{tabular}{cccccc}
\hline
$H$ & $W$ & \multicolumn{ 1 }{c}{ LA-S-1 } & \multicolumn{ 1 }{c}{ SM-2 } & \multicolumn{ 1 }{c}{ SmSEQ-2 } & \multicolumn{ 1 }{c}{ JZW } \\
\hline
10 & 10 &     0.00 &     0.00 &     0.00 &     0.00 \\
20 & 20 &     0.75 &     0.55 &     0.49 &     0.49 \\
30 & 30 &    12.74 &     8.47 &     7.00 &     7.33 \\
40 & 40 &   123.62 &    87.48 &    59.19 &    71.14 \\
50 & 50 &   769.24 &   515.80 &   329.88 &   434.28 \\
60 & 60 &  3579.65 &  2618.78 &  1509.08 &  2141.47 \\
70 & 70 & 13303.65 &  9031.78 &  4933.01 &  7534.28 \\
80 & 80 & 39538.48 & 27273.29 & 13638.50 & 21612.92 \\
90 & 90 & 113666.67 & 78958.13 & 35068.10 & 61040.70 \\
100 & 100 & 237755.00 (31) & 192664.73 (7) & 85194.16 & 154284.41 (3) \\
\hline
\end{tabular}
\caption{CPU time in seconds of LS, using different methods for the
  starting solution. Tricoire instances, $H_{max} = H+2$. The number
  of instances which could not be solved due to time
  restrictions is mentioned in parentheses.}
\label{tab:fast-large-H+2-CPU}
\end{table}

Information on the required CPU time in seconds 
for the instances of  \citet{tricoire2018new} 
can be found in Tables~\ref{tab:fast-large-unlimited-CPU}
and~\ref{tab:fast-large-H+2-CPU}. Each entry gives the average CPU
time across 40 instances per instance class and in parentheses the
number of instances which could not be solved due to CPU budget
limitation. Quite obviously, CPU times increase with instance size,
sometimes exceeding the time limit for the largest instances with $H_{max} =
H+2$. However, we note that for the largest instances the average obtained improvement amounts to at least 20\% for SM-2 and to more than 40\% when applied to solutions obtained by SmSEQ-2. We also note that these values are due to running the full fledged LS until termination. Truncated versions  with a reduced state-space or a CPU time limit may equally be envisaged.

For all other instances, the required CPU effort was systematically
less than a second; therefore we do not report it.

\section{Conclusions and perspectives}  \label{s:conclusion}

We have introduced the first local search based heuristic algorithm
for the unrestricted block relocation problem; it relies on dynamic
programming  and we were able to illustrate its value in extensive
tests on large benchmark instances, reaching improvements of up to
50\% compared to starting solutions. These starting solutions are
obtained by state-of-the-art constructive heuristics. Our results
nicely show the unexpectedly high potential for improvement on these
large BRP instances.

Future research will involve the design of more
efficient improvement methods, so as to obtain solutions of comparable
quality with lower time requirements. Apart from cases when the aspiration speedup is triggered, the local search aims at finding the best relocation sequence; it potentially could be accelerated by limiting the search to any improving sequence or by voluntarily limiting the number of states explored. In addition, the local search operator could be integrated into a metaheuristic scheme, instead of being used in a simple greedy descent algorithm.

Another perspective of our work is the design of more complex local search schemes which do not only consider a single container at a time, but two or more. Our results for the wide instances indicate the need for such more complex operators.

Extensions of the dynamic programming scheme to other variants of the BRP will also be investigated. A natural extension is to the restricted BRP. Less transitions would be accepted, which would reduce the state-space and accelerate the algorithm, but it would also limit the opportunities for improvement. 
Problems in which the priority order is defined among groups of containers could also be considered, with the opposite effect: the state-space would increase a lot, so as the chances of improvement.

\section*{Acknowledgments}

The authors thank the referees for their acute comments, which significantly contributed to enhance the quality of this paper.

\bibliographystyle{apalike}
\bibliography{crp}

\begin{appendix}
\section{Results for Caserta et al. instances}
Tables~\ref{tab:advanced-caserta-unlimited}--\ref{tab:fast-caserta-H+2}
provide the following information for each instance class
from~\citet{caserta2011applying}: the average improvement
(Avg. impr.) obtained by means of LS, the number of instances for
which LS found an improved solution (\# impr.) and the average CPU
effort of LS in seconds.
Average improvement represents the average number of
relocations saved by applying local search.
These indicators are
reported for each of the four
advanced methods (rake search and pilot method
of~\citet{tricoire2018new}, two versions of the greedy look ahead
heuristic (GLAH) of~\citet{jin2015solving}) and for the four fast
constructive methods (LA-S-1, SM-2, SmSEQ-2, JZW). Each row represents
40 instances.

\begin{table}
\scriptsize
\centering
\begin{tabular}{cccccccccccccc}
\hline
& & \multicolumn{ 3 }{c}{ Rake search } & \multicolumn{ 3 }{c}{ Pilot method } & \multicolumn{ 3 }{c}{ GLAH-3 } & \multicolumn{ 3 }{c}{ GLAH-4 } \\
& & Avg. & \#  &  & Avg. & \#  & & Avg.  & \#  &  & Avg.  & \#  &  \\
$H$ & $W$ & impr. & impr. & CPU &  impr. &  impr. & CPU &  impr. & impr. & CPU & impr. &  impr. & CPU \\
\hline
 3 &  3 &     0.00 &        0 &     0.00 &     0.00 &        0 &     0.00 &     0.00 &        0 &     0.00 &     0.00 &        0 &     0.00 \\
 3 &  4 &     0.00 &        0 &     0.00 &     0.00 &        0 &     0.00 &     0.00 &        0 &     0.00 &     0.00 &        0 &     0.00 \\
 3 &  5 &     0.00 &        0 &     0.00 &     0.00 &        0 &     0.00 &     0.00 &        0 &     0.00 &     0.00 &        0 &     0.00 \\
 3 &  6 &     0.00 &        0 &     0.00 &     0.00 &        0 &     0.00 &     0.00 &        0 &     0.00 &     0.00 &        0 &     0.00 \\
 3 &  7 &     0.00 &        0 &     0.00 &     0.00 &        0 &     0.00 &     0.00 &        0 &     0.00 &     0.00 &        0 &     0.00 \\
 3 &  8 &     0.00 &        0 &     0.00 &     0.00 &        0 &     0.00 &     0.00 &        0 &     0.00 &     0.00 &        0 &     0.00 \\
 4 &  4 &     0.00 &        0 &     0.00 &     0.00 &        0 &     0.00 &     0.00 &        0 &     0.00 &     0.00 &        0 &     0.00 \\
 4 &  5 &     0.00 &        0 &     0.00 &     0.00 &        0 &     0.00 &     0.03 &        1 &     0.00 &     0.03 &        1 &     0.00 \\
 4 &  6 &     0.00 &        0 &     0.00 &     0.00 &        0 &     0.00 &     0.00 &        0 &     0.00 &     0.00 &        0 &     0.00 \\
 4 &  7 &     0.00 &        0 &     0.00 &     0.00 &        0 &     0.00 &     0.00 &        0 &     0.00 &     0.00 &        0 &     0.00 \\
 5 &  4 &     0.00 &        0 &     0.00 &     0.03 &        1 &     0.00 &     0.00 &        0 &     0.00 &     0.00 &        0 &     0.00 \\
 5 &  5 &     0.00 &        0 &     0.00 &     0.03 &        1 &     0.00 &     0.00 &        0 &     0.00 &     0.00 &        0 &     0.00 \\
 5 &  6 &     0.00 &        0 &     0.00 &     0.00 &        0 &     0.00 &     0.00 &        0 &     0.00 &     0.00 &        0 &     0.00 \\
 5 &  7 &     0.00 &        0 &     0.00 &     0.00 &        0 &     0.00 &     0.03 &        1 &     0.00 &     0.00 &        0 &     0.00 \\
 5 &  8 &     0.00 &        0 &     0.00 &     0.00 &        0 &     0.00 &     0.00 &        0 &     0.00 &     0.00 &        0 &     0.00 \\
 5 &  9 &     0.00 &        0 &     0.00 &     0.03 &        1 &     0.00 &     0.00 &        0 &     0.00 &     0.00 &        0 &     0.00 \\
 5 & 10 &     0.03 &        1 &     0.00 &     0.00 &        0 &     0.00 &     0.00 &        0 &     0.00 &     0.00 &        0 &     0.00 \\
 6 &  6 &     0.07 &        3 &     0.00 &     0.07 &        3 &     0.00 &     0.00 &        0 &     0.00 &     0.00 &        0 &     0.00 \\
 6 & 10 &     0.03 &        1 &     0.00 &     0.03 &        1 &     0.00 &     0.00 &        0 &     0.00 &     0.03 &        1 &     0.00 \\
10 &  6 &     2.58 &       35 &     0.00 &     1.80 &       30 &     0.00 &     1.15 &       24 &     0.00 &     1.27 &       29 &     0.00 \\
10 & 10 &     1.55 &       29 &     0.00 &     0.72 &       20 &     0.00 &     0.90 &       21 &     0.00 &     0.82 &       18 &     0.00 \\
\hline
\end{tabular}
\caption{Improvement by LS, using different methods for the starting solution. Caserta instances, $H_{max} = unlimited$.}
\label{tab:advanced-caserta-unlimited}
\end{table}
\begin{table}
\scriptsize
\centering
\begin{tabular}{cccccccccccccc}
\hline
& & \multicolumn{ 3 }{c}{ Rake search } & \multicolumn{ 3 }{c}{ Pilot method } & \multicolumn{ 3 }{c}{ GLAH-3 } & \multicolumn{ 3 }{c}{ GLAH-4 } \\
& & Avg. & \# &  & Avg.  & \#  & & Avg.  & \#  & & Avg.  & \#  &  \\
$H$ & $W$ & impr. &  impr. & CPU &  impr. & impr. & CPU &  impr. &  impr. & CPU & impr. &  impr. & CPU \\
\hline
 3 &  3 &     0.00 &        0 &     0.00 &     0.00 &        0 &     0.00 &     0.00 &        0 &     0.00 &     0.00 &        0 &     0.00 \\
 3 &  4 &     0.00 &        0 &     0.00 &     0.00 &        0 &     0.00 &     0.00 &        0 &     0.00 &     0.00 &        0 &     0.00 \\
 3 &  5 &     0.00 &        0 &     0.00 &     0.00 &        0 &     0.00 &     0.00 &        0 &     0.00 &     0.00 &        0 &     0.00 \\
 3 &  6 &     0.00 &        0 &     0.00 &     0.00 &        0 &     0.00 &     0.00 &        0 &     0.00 &     0.00 &        0 &     0.00 \\
 3 &  7 &     0.00 &        0 &     0.00 &     0.00 &        0 &     0.00 &     0.00 &        0 &     0.00 &     0.00 &        0 &     0.00 \\
 3 &  8 &     0.00 &        0 &     0.00 &     0.00 &        0 &     0.00 &     0.00 &        0 &     0.00 &     0.00 &        0 &     0.00 \\
 4 &  4 &     0.00 &        0 &     0.00 &     0.00 &        0 &     0.00 &     0.00 &        0 &     0.00 &     0.00 &        0 &     0.00 \\
 4 &  5 &     0.00 &        0 &     0.00 &     0.00 &        0 &     0.00 &     0.03 &        1 &     0.00 &     0.03 &        1 &     0.00 \\
 4 &  6 &     0.00 &        0 &     0.00 &     0.00 &        0 &     0.00 &     0.00 &        0 &     0.00 &     0.00 &        0 &     0.00 \\
 4 &  7 &     0.00 &        0 &     0.00 &     0.00 &        0 &     0.00 &     0.00 &        0 &     0.00 &     0.00 &        0 &     0.00 \\
 5 &  4 &     0.03 &        1 &     0.00 &     0.03 &        1 &     0.00 &     0.05 &        2 &     0.00 &     0.03 &        1 &     0.00 \\
 5 &  5 &     0.00 &        0 &     0.00 &     0.03 &        1 &     0.00 &     0.03 &        1 &     0.00 &     0.00 &        0 &     0.00 \\
 5 &  6 &     0.00 &        0 &     0.00 &     0.00 &        0 &     0.00 &     0.00 &        0 &     0.00 &     0.00 &        0 &     0.00 \\
 5 &  7 &     0.03 &        1 &     0.00 &     0.00 &        0 &     0.00 &     0.03 &        1 &     0.00 &     0.03 &        1 &     0.00 \\
 5 &  8 &     0.00 &        0 &     0.00 &     0.00 &        0 &     0.00 &     0.03 &        1 &     0.00 &     0.00 &        0 &     0.00 \\
 5 &  9 &     0.03 &        1 &     0.00 &     0.03 &        1 &     0.00 &     0.00 &        0 &     0.00 &     0.00 &        0 &     0.00 \\
 5 & 10 &     0.00 &        0 &     0.00 &     0.00 &        0 &     0.00 &     0.00 &        0 &     0.00 &     0.00 &        0 &     0.00 \\
 6 &  6 &     0.07 &        2 &     0.00 &     0.05 &        2 &     0.00 &     0.00 &        0 &     0.00 &     0.00 &        0 &     0.00 \\
 6 & 10 &     0.05 &        2 &     0.00 &     0.03 &        1 &     0.00 &     0.00 &        0 &     0.00 &     0.00 &        0 &     0.00 \\
10 &  6 &     1.50 &       25 &     0.00 &     0.95 &       20 &     0.00 &     0.88 &       17 &     0.00 &     0.88 &       21 &     0.00 \\
10 & 10 &     1.00 &       21 &     0.00 &     0.47 &       16 &     0.00 &     0.53 &       13 &     0.00 &     0.35 &       11 &     0.00 \\
\hline
\end{tabular}
\caption{Improvement by LS, using different methods for the starting solution. Caserta instances, $H_{max} = H+2$.}
\label{tab:advanced-caserta-H+2}
\end{table}

\begin{table}
\scriptsize
\centering
\begin{tabular}{cccccccccccccc}
\hline
& & \multicolumn{ 3 }{c}{ LA-S-1 } & \multicolumn{ 3 }{c}{ SM-2 } & \multicolumn{ 3 }{c}{ SmSEQ-2 } & \multicolumn{ 3 }{c}{ JZW } \\
 & & Avg. & \# &  & Avg.  & \#  & & Avg.  & \#  & & Avg.  & \#  &  \\
$H$ & $W$ & impr. &  impr. & CPU &  impr. & impr. & CPU &  impr. &  impr. & CPU & impr. &  impr. & CPU \\
\hline
 3 &  3 &     0.03 &        1 &     0.00 &     0.00 &        0 &     0.00 &     0.00 &        0 &     0.00 &     0.03 &        1 &     0.00 \\
 3 &  4 &     0.00 &        0 &     0.00 &     0.00 &        0 &     0.00 &     0.00 &        0 &     0.00 &     0.00 &        0 &     0.00 \\
 3 &  5 &     0.00 &        0 &     0.00 &     0.00 &        0 &     0.00 &     0.00 &        0 &     0.00 &     0.00 &        0 &     0.00 \\
 3 &  6 &     0.00 &        0 &     0.00 &     0.00 &        0 &     0.00 &     0.00 &        0 &     0.00 &     0.00 &        0 &     0.00 \\
 3 &  7 &     0.00 &        0 &     0.00 &     0.00 &        0 &     0.00 &     0.00 &        0 &     0.00 &     0.00 &        0 &     0.00 \\
 3 &  8 &     0.00 &        0 &     0.00 &     0.00 &        0 &     0.00 &     0.00 &        0 &     0.00 &     0.00 &        0 &     0.00 \\
 4 &  4 &     0.25 &        8 &     0.00 &     0.07 &        3 &     0.00 &     0.12 &        4 &     0.00 &     0.03 &        1 &     0.00 \\
 4 &  5 &     0.07 &        3 &     0.00 &     0.00 &        0 &     0.00 &     0.05 &        2 &     0.00 &     0.03 &        1 &     0.00 \\
 4 &  6 &     0.07 &        2 &     0.00 &     0.00 &        0 &     0.00 &     0.03 &        1 &     0.00 &     0.07 &        3 &     0.00 \\
 4 &  7 &     0.03 &        1 &     0.00 &     0.03 &        1 &     0.00 &     0.03 &        1 &     0.00 &     0.03 &        1 &     0.00 \\
 5 &  4 &     0.45 &       14 &     0.00 &     0.33 &        7 &     0.00 &     0.30 &       10 &     0.00 &     0.20 &        7 &     0.00 \\
 5 &  5 &     0.57 &       20 &     0.00 &     0.23 &        6 &     0.00 &     0.53 &       15 &     0.00 &     0.35 &       13 &     0.00 \\
 5 &  6 &     0.25 &        8 &     0.00 &     0.05 &        2 &     0.00 &     0.17 &        5 &     0.00 &     0.12 &        5 &     0.00 \\
 5 &  7 &     0.30 &        7 &     0.00 &     0.00 &        0 &     0.00 &     0.20 &        6 &     0.00 &     0.17 &        7 &     0.00 \\
 5 &  8 &     0.17 &        5 &     0.00 &     0.00 &        0 &     0.00 &     0.05 &        2 &     0.00 &     0.05 &        1 &     0.00 \\
 5 &  9 &     0.12 &        5 &     0.00 &     0.00 &        0 &     0.00 &     0.17 &        5 &     0.00 &     0.07 &        2 &     0.00 \\
 5 & 10 &     0.05 &        2 &     0.00 &     0.00 &        0 &     0.00 &     0.15 &        5 &     0.00 &     0.05 &        2 &     0.00 \\
 6 &  6 &     1.23 &       22 &     0.00 &     0.53 &       15 &     0.00 &     0.80 &       19 &     0.00 &     0.50 &       15 &     0.00 \\
 6 & 10 &     0.42 &       11 &     0.00 &     0.05 &        1 &     0.00 &     0.28 &        6 &     0.00 &     0.20 &        7 &     0.00 \\
10 &  6 &    10.22 &       40 &     0.00 &     8.00 &       40 &     0.00 &     9.07 &       40 &     0.00 &     6.80 &       39 &     0.00 \\
10 & 10 &    11.65 &       39 &     0.00 &     4.20 &       37 &     0.00 &     6.20 &       39 &     0.00 &     3.90 &       39 &     0.00 \\
\hline
\end{tabular}
\caption{Improvement by LS, using different methods for the starting solution. Caserta instances, $H_{max} = unlimited$.}
\label{tab:fast-caserta-unlimited}
\end{table}
\begin{table}
\scriptsize
\centering
\begin{tabular}{cccccccccccccc}
\hline
& & \multicolumn{ 3 }{c}{ LA-S-1 } & \multicolumn{ 3 }{c}{ SM-2 } & \multicolumn{ 3 }{c}{ SmSEQ-2 } & \multicolumn{ 3 }{c}{ JZW } \\
 & & Avg. & \# &  & Avg.  & \#  & & Avg.  & \#  & & Avg.  & \#  &  \\
$H$ & $W$ & impr. &  impr. & CPU &  impr. & impr. & CPU &  impr. &  impr. & CPU & impr. &  impr. & CPU \\
\hline
 3 &  3 &     0.03 &        1 &     0.00 &     0.00 &        0 &     0.00 &     0.00 &        0 &     0.00 &     0.03 &        1 &     0.00 \\
 3 &  4 &     0.00 &        0 &     0.00 &     0.00 &        0 &     0.00 &     0.00 &        0 &     0.00 &     0.00 &        0 &     0.00 \\
 3 &  5 &     0.00 &        0 &     0.00 &     0.00 &        0 &     0.00 &     0.00 &        0 &     0.00 &     0.00 &        0 &     0.00 \\
 3 &  6 &     0.00 &        0 &     0.00 &     0.00 &        0 &     0.00 &     0.00 &        0 &     0.00 &     0.00 &        0 &     0.00 \\
 3 &  7 &     0.00 &        0 &     0.00 &     0.00 &        0 &     0.00 &     0.00 &        0 &     0.00 &     0.00 &        0 &     0.00 \\
 3 &  8 &     0.00 &        0 &     0.00 &     0.00 &        0 &     0.00 &     0.00 &        0 &     0.00 &     0.00 &        0 &     0.00 \\
 4 &  4 &     0.20 &        7 &     0.00 &     0.00 &        0 &     0.00 &     0.05 &        2 &     0.00 &     0.05 &        2 &     0.00 \\
 4 &  5 &     0.10 &        4 &     0.00 &     0.00 &        0 &     0.00 &     0.07 &        3 &     0.00 &     0.05 &        2 &     0.00 \\
 4 &  6 &     0.10 &        3 &     0.00 &     0.00 &        0 &     0.00 &     0.03 &        1 &     0.00 &     0.05 &        2 &     0.00 \\
 4 &  7 &     0.03 &        1 &     0.00 &     0.00 &        0 &     0.00 &     0.03 &        1 &     0.00 &     0.07 &        3 &     0.00 \\
 5 &  4 &     0.40 &       11 &     0.00 &     0.20 &        7 &     0.00 &     0.33 &        9 &     0.00 &     0.28 &        9 &     0.00 \\
 5 &  5 &     0.53 &       14 &     0.00 &     0.07 &        3 &     0.00 &     0.45 &       11 &     0.00 &     0.15 &        6 &     0.00 \\
 5 &  6 &     0.12 &        4 &     0.00 &     0.07 &        2 &     0.00 &     0.17 &        5 &     0.00 &     0.12 &        5 &     0.00 \\
 5 &  7 &     0.17 &        5 &     0.00 &     0.00 &        0 &     0.00 &     0.23 &        6 &     0.00 &     0.15 &        6 &     0.00 \\
 5 &  8 &     0.12 &        4 &     0.00 &     0.03 &        1 &     0.00 &     0.07 &        3 &     0.00 &     0.12 &        4 &     0.00 \\
 5 &  9 &     0.15 &        5 &     0.00 &     0.07 &        2 &     0.00 &     0.25 &       10 &     0.00 &     0.07 &        3 &     0.00 \\
 5 & 10 &     0.03 &        1 &     0.00 &     0.03 &        1 &     0.00 &     0.17 &        5 &     0.00 &     0.10 &        4 &     0.00 \\
 6 &  6 &     0.90 &       23 &     0.00 &     0.30 &       11 &     0.00 &     0.88 &       18 &     0.00 &     0.55 &       15 &     0.00 \\
 6 & 10 &     0.35 &       10 &     0.00 &     0.03 &        1 &     0.00 &     0.30 &        8 &     0.00 &     0.25 &        8 &     0.00 \\
10 &  6 &     7.17 &       39 &     0.00 &     5.62 &       38 &     0.00 &     6.88 &       39 &     0.00 &     4.33 &       36 &     0.00 \\
10 & 10 &     7.58 &       37 &     0.00 &     2.65 &       33 &     0.00 &     5.78 &       40 &     0.00 &     3.02 &       35 &     0.00 \\
\hline
\end{tabular}
\caption{Improvement by LS, using different methods for the starting solution. Caserta instances, $H_{max} = H+2$.}
\label{tab:fast-caserta-H+2}
\end{table}

\section{Detailed results for Tricoire instances}
Tables~\ref{tab:fast-large-avg-unlimited} and
\ref{tab:fast-large-avg-H+2} provide average values across 40
instances per instance class of the instances
from~\citet{tricoire2018new}.
The starting solutions are obtained by means of
the fast constructive methods, then LS is applied. Average values
before and after LS are reported; in addition,
the percentage gap between these two values is provided.
Percentage gap is computed as $100 \frac{b-a}{b}$, where $b$ is the number of
relocations before local search and $a$ is the number of relocations
after local search.
In the case
where not all of the 40 instances belonging to an instance class could
be solved, the value in parentheses gives the number of instance
which were not solved due to time restrictions. This only
happens for the largest instances with $H_{max} = H+2$.
\begin{landscape}
\begin{table}
\scriptsize
\centering
\begin{tabular}{cccccccccccccc}
\hline
& & \multicolumn{ 3 }{c}{ LA-S-1 } & \multicolumn{ 3 }{c}{ SM-2 } & \multicolumn{ 3 }{c}{ SmSEQ-2 } & \multicolumn{ 3 }{c}{ JZW } \\
$H$ & $W$ & before LS & after LS & \% gap & before LS & after LS & \% gap & before LS & after LS & \% gap & before LS & after LS & \% gap \\
\hline
10 & 10 &   117.78 &   107.55 &     8.45 &   109.47 &   104.97 &     4.00 &   111.17 &   105.17 &     5.29 &   109.45 &   104.67 &     4.27 \\
20 & 20 &   688.75 &   548.10 &    20.33 &   623.33 &   531.02 &    14.68 &   616.73 &   525.80 &    14.67 &   610.20 &   528.17 &    13.35 \\
30 & 30 &  1927.45 &  1367.22 &    28.93 &  1770.45 &  1323.97 &    25.15 &  1720.67 &  1315.08 &    23.51 &  1700.65 &  1319.10 &    22.31 \\
40 & 40 &  4025.45 &  2602.43 &    35.30 &  3656.85 &  2524.82 &    30.92 &  3520.90 &  2457.88 &    30.09 &  3488.72 &  2520.03 &    27.73 \\
50 & 50 &  7104.00 &  4256.65 &    40.03 &  6520.48 &  4113.25 &    36.89 &  6238.27 &  3982.88 &    36.09 &  6175.85 &  4113.05 &    33.37 \\
60 & 60 & 11295.30 &  6388.70 &    43.39 & 10377.52 &  6174.18 &    40.46 &  9622.20 &  5805.75 &    39.60 & 10007.05 &  6218.82 &    37.81 \\
70 & 70 & 16683.50 &  8932.38 &    46.40 & 15501.00 &  8684.42 &    43.95 & 14316.00 &  8070.27 &    43.57 & 14935.65 &  8671.67 &    41.90 \\
80 & 80 & 23121.35 & 11802.95 &    48.88 & 21620.28 & 11512.45 &    46.73 & 19659.33 & 10605.10 &    46.01 & 20867.50 & 11533.15 &    44.69 \\
90 & 90 & 31189.08 & 15241.38 &    51.10 & 29429.58 & 14858.75 &    49.50 & 25889.85 & 13558.40 &    47.59 & 28114.05 & 14872.77 &    47.08 \\
100 & 100 & 40199.30 & 19191.75 &    52.21 & 37911.85 & 18616.22 &    50.87 & 33403.68 & 16857.03 &    49.51 & 36567.03 & 18730.12 &    48.74 \\
\hline
\end{tabular}
\caption{Average objective values and improvement by LS, using different methods for the starting solution. Large instances, $H_{max} = unlimited$.}
\label{tab:fast-large-avg-unlimited}
\end{table}

\begin{table}
\scriptsize
\centering
\begin{tabular}{cccccccccccccc}
\hline
& & \multicolumn{ 3 }{c}{ LA-S-1 } & \multicolumn{ 3 }{c}{ SM-2 } & \multicolumn{ 3 }{c}{ SmSEQ-2 } & \multicolumn{ 3 }{c}{ JZW } \\
$H$ & $W$ & before LS & after LS & \% gap & before LS & after LS & \% gap & before LS & after LS & \% gap & before LS & after LS & \% gap \\
\hline
10 & 10 &   125.00 &   119.85 &     4.07 &   117.17 &   114.28 &     2.44 &   121.62 &   115.28 &     5.01 &   117.67 &   114.72 &     2.47 \\
20 & 20 &   825.02 &   715.50 &    13.23 &   735.83 &   681.67 &     7.32 &   735.12 &   625.30 &    14.81 &   722.20 &   675.05 &     6.50 \\
30 & 30 &  2552.22 &  2070.05 &    18.86 &  2240.97 &  1969.72 &    12.11 &  2185.97 &  1700.60 &    22.15 &  2179.10 &  1933.95 &    11.22 \\
40 & 40 &  5885.30 &  4437.35 &    24.59 &  5035.05 &  4196.85 &    16.62 &  4861.10 &  3493.85 &    28.05 &  4908.20 &  4119.52 &    16.04 \\
50 & 50 & 11288.48 &  8040.82 &    28.76 &  9497.15 &  7591.98 &    20.07 &  9046.42 &  6118.60 &    32.33 &  9346.12 &  7422.10 &    20.58 \\
60 & 60 & 19320.25 & 13308.52 &    31.11 & 16046.17 & 12501.33 &    22.08 & 15349.50 &  9754.60 &    36.35 & 16121.12 & 12277.48 &    23.83 \\
70 & 70 & 30484.30 & 20502.67 &    32.74 & 25106.08 & 19082.20 &    23.99 & 23513.88 & 14386.62 &    38.73 & 25198.92 & 18693.78 &    25.81 \\
80 & 80 & 45335.72 & 29872.88 &    34.13 & 36906.50 & 27348.20 &    25.89 & 34368.60 & 20078.97 &    41.48 & 37639.85 & 27009.28 &    28.24 \\
90 & 90 & 64057.75 & 41523.95 &    35.18 & 51892.47 & 37706.82 &    27.33 & 47476.97 & 26978.70 &    43.06 & 53131.47 & 37191.57 &    30.00 \\
100 & 100 & 86280.33 & 55446.56 &    35.73 (31) & 70045.58 & 50571.85 &    27.81 (7) & 63864.85 & 35418.05 &    44.47 & 73151.03 & 50730.54 &    30.66 (3) \\
\hline
\end{tabular}
\caption{Average objective values and improvement by LS, using different methods for the starting solution. Large instances, $H_{max} = H+2$.}
\label{tab:fast-large-avg-H+2}
\end{table}
\end{landscape}

\section{Detailed results for wide instances}

As in the previous appendix, we report average objective values before
and after LS as well as the percentage gap between these two values,
this time for wide instances. Tables~\ref{tab:fast-wide-avg-unlimited}
and~\ref{tab:fast-wide-avg-H+2} present these results. 

\begin{landscape}
\begin{table}
\scriptsize
\centering
\begin{tabular}{cccccccccccccc}
\hline
& & \multicolumn{ 3 }{c}{ LA-S-1 } & \multicolumn{ 3 }{c}{ SM-2 } & \multicolumn{ 3 }{c}{ SmSEQ-2 } & \multicolumn{ 3 }{c}{ JZW } \\
$H$ & $W$ & before LS & after LS & \% gap & before LS & after LS & \% gap & before LS & after LS & \% gap & before LS & after LS & \% gap \\
\hline
10 & 10 &   115.65 &   105.55 &     8.53 &   108.25 &   103.20 &     4.53 &   107.38 &   102.47 &     4.48 &   107.10 &   102.17 &     4.49 \\
10 & 20 &   190.38 &   184.47 &     3.05 &   176.28 &   175.43 &     0.47 &   178.72 &   175.43 &     1.81 &   179.75 &   178.25 &     0.81 \\
10 & 30 &   262.18 &   259.45 &     1.04 &   244.12 &   244.03 &     0.04 &   247.00 &   245.03 &     0.79 &   251.20 &   250.93 &     0.11 \\
10 & 40 &   336.77 &   334.82 &     0.57 &   316.00 &   315.90 &     0.03 &   319.95 &   316.60 &     1.03 &   324.85 &   324.75 &     0.03 \\
10 & 50 &   408.68 &   407.90 &     0.19 &   384.25 &   384.25 &     0.00 &   388.55 &   386.02 &     0.64 &   397.23 &   397.23 &     0.00 \\
10 & 60 &   476.62 &   476.10 &     0.11 &   452.50 &   452.38 &     0.03 &   456.43 &   453.75 &     0.58 &   468.25 &   468.23 &     0.01 \\
10 & 70 &   549.42 &   549.15 &     0.05 &   522.12 &   522.08 &     0.01 &   527.02 &   524.60 &     0.46 &   540.95 &   540.95 &     0.00 \\
10 & 80 &   619.83 &   619.65 &     0.03 &   589.70 &   589.60 &     0.02 &   595.55 &   593.23 &     0.39 &   611.50 &   611.48 &     0.00 \\
10 & 90 &   692.85 &   692.73 &     0.02 &   665.48 &   665.48 &     0.00 &   669.98 &   667.25 &     0.40 &   685.38 &   685.38 &     0.00 \\
10 & 100 &   765.12 &   765.05 &     0.01 &   734.67 &   734.65 &     0.00 &   739.17 &   737.08 &     0.28 &   758.95 &   758.92 &     0.00 \\
\hline
\end{tabular}
\caption{Average objective values and improvement by LS, using different methods for the starting solution. Wide instances, $H_{max} = unlimited$.}
\label{tab:fast-wide-avg-unlimited}
\end{table}
\begin{table}
\scriptsize
\centering
\begin{tabular}{cccccccccccccc}
\hline
& & \multicolumn{ 3 }{c}{ LA-S-1 } & \multicolumn{ 3 }{c}{ SM-2 } & \multicolumn{ 3 }{c}{ SmSEQ-2 } & \multicolumn{ 3 }{c}{ JZW } \\
$H$ & $W$ & before LS & after LS & \% gap & before LS & after LS & \% gap & before LS & after LS & \% gap & before LS & after LS & \% gap \\
\hline
10 & 10 &   125.28 &   117.83 &     5.79 &   115.20 &   113.05 &     1.82 &   118.58 &   112.12 &     5.29 &   114.88 &   112.00 &     2.49 \\
10 & 20 &   207.47 &   203.57 &     1.87 &   195.55 &   194.05 &     0.75 &   195.25 &   191.50 &     1.87 &   196.80 &   195.12 &     0.85 \\
10 & 30 &   288.20 &   286.07 &     0.74 &   273.23 &   272.70 &     0.18 &   273.88 &   269.77 &     1.48 &   278.43 &   277.38 &     0.38 \\
10 & 40 &   370.77 &   368.73 &     0.55 &   351.90 &   351.38 &     0.15 &   351.48 &   347.27 &     1.18 &   358.30 &   357.30 &     0.28 \\
10 & 50 &   453.45 &   452.25 &     0.27 &   434.07 &   433.70 &     0.09 &   430.20 &   426.60 &     0.82 &   442.85 &   441.77 &     0.24 \\
10 & 60 &   531.67 &   530.65 &     0.19 &   509.50 &   508.90 &     0.12 &   502.60 &   498.50 &     0.81 &   522.10 &   521.38 &     0.14 \\
10 & 70 &   614.25 &   613.17 &     0.17 &   586.02 &   585.38 &     0.11 &   581.90 &   577.12 &     0.81 &   604.48 &   603.67 &     0.13 \\
10 & 80 &   705.02 &   704.05 &     0.14 &   668.62 &   668.02 &     0.09 &   663.98 &   659.42 &     0.68 &   687.33 &   686.38 &     0.14 \\
10 & 90 &   781.77 &   781.33 &     0.06 &   745.15 &   744.60 &     0.07 &   740.73 &   735.08 &     0.76 &   772.05 &   771.27 &     0.10 \\
10 & 100 &   865.75 &   865.33 &     0.05 &   822.20 &   821.42 &     0.09 &   821.98 &   816.98 &     0.60 &   855.10 &   854.10 &     0.12 \\
\hline
\end{tabular}
\caption{Average objective values and improvement by LS, using different methods for the starting solution. Wide instances, $H_{max} = H+2$.}
\label{tab:fast-wide-avg-H+2}
\end{table}

\end{landscape}

\end{appendix}

\end{document}